\documentclass{emulateapj}
\usepackage{amsmath,amstext,amssymb}
\usepackage{float}
\usepackage{multirow}
\usepackage{relsize}
\usepackage{natbib}

\usepackage[dvipsnames]{xcolor}
\usepackage{hyperref}
\usepackage{cleveref}

\newcommand\myshade{85}
\colorlet{mylinkcolor}{violet}
\colorlet{mycitecolor}{Blue}
\colorlet{myurlcolor}{Aquamarine}

\hypersetup{
  linkcolor  = mylinkcolor!\myshade!black,
  citecolor  = mycitecolor!\myshade!black,
  urlcolor   = myurlcolor!\myshade!black,
  colorlinks = true,
}

\DeclareSymbolFont{matha}{OML}{txmi1}{m}{it}
\DeclareMathSymbol{\varv}{\mathord}{matha}{118}

\defcitealias{Sternberg2014}{S14}

\newcommand{\Ma}{\mathcal{M}_{\rm A}}
\newcommand{\Ms}{\mathcal{M}_{\rm s}}
\newcommand{\cs}{c_{\rm s}}
\newcommand{\Nb}{N_{\rm b}}

\def\kms{km~s$^{-1}$}
\def\ga{\mathrel{\hbox{\rlap{\hbox{\lower4pt\hbox{$\sim$}}}\hbox{$>$}}}}
\def\la{\mathrel{\hbox{\rlap{\hbox{\lower4pt\hbox{$\sim$}}}\hbox{$<$}}}}

\usepackage{scalerel,tikz}
\usetikzlibrary{svg.path}
\definecolor{orcidlogocol}{HTML}{A6CE39}
\tikzset{orcidlogo/.pic={
 \fill[orcidlogocol] svg{M256,128c0,70.7-57.3,128-128,128C57.3,256,0,198.7,0,128C0,57.3,57.3,0,128,0C198.7,0,256,57.3,256,128z};
 \fill[white] svg{M86.3,186.2H70.9V79.1h15.4v48.4V186.2z}
 svg{M108.9,79.1h41.6c39.6,0,57,28.3,57,53.6c0,27.5-21.5,53.6-56.8,53.6h-41.8V79.1z M124.3,172.4h24.5c34.9,0,42.9-26.5,42.9-39.7c0-21.5-13.7-39.7-43.7-39.7h-23.7V172.4z}
 svg{M88.7,56.8c0,5.5-4.5,10.1-10.1,10.1c-5.6,0-10.1-4.6-10.1-10.1c0-5.6,4.5-10.1,10.1-10.1C84.2,46.7,88.7,51.3,88.7,56.8z};
}}
\newcommand\orcidicon[1]{\href{https://orcid.org/#1}{\mbox{\scalerel*{
\begin{tikzpicture}[yscale=-1,transform shape]
\pic{orcidlogo};
\end{tikzpicture}
}{|}}}}

\newcommand{\despotic}{\texttt{DESPOTIC}}

\shorttitle{The Catalogue for Astrophysical MHD Turbulence Simulations (CATS)}

\shortauthors{Burkhart et al.}
\begin{document}
\title{The Catalogue for Astrophysical Turbulence Simulations (CATS)}

\email{Corresponding Author: Blakesley Burkhart \\
b.burkhart@physics.rutgers.edu}

\author{Burkhart, B.\altaffilmark{1,2}~\orcidicon{0000-0001-5817-5944},
Appel, S. M.\altaffilmark{1},
Bialy, S.\altaffilmark{3}~\orcidicon{0000-0002-0404-003X}, 
Cho, J.\altaffilmark{4},
Christensen, A. J.\altaffilmark{5},
Collins, D.\altaffilmark{6}, 
Federrath, C.\altaffilmark{7,10}~\orcidicon{0000-0002-0706-2306},
Fielding, D.\altaffilmark{2}~\orcidicon{0000-0003-3806-8548},
Finkbeiner, D.\altaffilmark{3}, 
Hill, A. S.\altaffilmark{8,9}~\orcidicon{0000-0001-7301-5666},
Ib\'a\~nez-Mej\'{\i}a, J. C.\altaffilmark{13}~\orcidicon{0000-0002-9868-3561},
Krumholz, M. R.\altaffilmark{7,10}~\orcidicon{0000-0003-3893-854X},
Lazarian, A.\altaffilmark{11,2},
Li, M. \altaffilmark{2} ~\orcidicon{0000-0003-0773-582X},
Mocz, P.\altaffilmark{12}~\orcidicon{0000-0001-6631-2566},
Mac Low, M.-M.\altaffilmark{2,13}~\orcidicon{0000-0003-0064-4060},
Naiman, J.\altaffilmark{14}~\orcidicon{0000-0002-9397-6189}, 
Portillo, S. K. N.\altaffilmark{15}~\orcidicon{0000-0001-8132-8056
}, 
Shane, B.\altaffilmark{1},
Slepian, Z.\altaffilmark{16}, 
Yuan, Y.\altaffilmark{7}~\orcidicon{0000-0001-6816-0682} }

\altaffiltext{1}{Department of Physics and Astronomy, Rutgers University,  136 Frelinghuysen Rd, Piscataway, NJ 08854, USA}
\altaffiltext{2}{Center for Computational Astrophysics, Flatiron Institute, 162 Fifth Avenue, New York, NY 10010, USA}
\altaffiltext{3}{Harvard-Smithsonian Center for Astrophysics, 60 Garden st. Cambridge, Ma, USA}
\altaffiltext{4}{Chungnam National University, Daejeon 34134, Republic of Korea}
\altaffiltext{5}{National Center for Supercomputing Applications, University of Illinois,  1205 W Clark St, Urbana, IL 61801, USA }
\altaffiltext{6}{Department of Physics, Florida State University, Tallahassee, FL 32306-4350, USA}
\altaffiltext{7}{Research School of Astronomy and Astrophysics, Australian National University, Canberra, ACT~2611, Australia}
\altaffiltext{8}{Department of Computer Science, Math, Physics, \& Statistics, Irving K. Barber Faculty of Science, University of British Columbia, Okanagan Campus, Kelowna, BC V1V 1V7 Canada}
\altaffiltext{9}{Dominion Radio Astrophysical Observatory, Herzberg Research Centre for Astronomy \& Astrophysics, National Research Council Canada, Penticton, BC V2A 6J9 Canada}
\altaffiltext{10}{ARC Centre of Excellence for Astronomy in Three Dimensions (ASTRO-3D), Canberra, ACT~2611, Australia}
\altaffiltext{11}{Astronomy Department, University of Wisconsin, Madison, WI 53711, USA}
\altaffiltext{12}{Einstein Fellow, Dept. of Astrophysical Sciences, 4 Ivy Lane, Princeton University, Princeton, NJ 08544, USA }
\altaffiltext{13}{Department of Astrophysics, American Museum of Natural History, 79th St at Central Park West, New York, NY 10024, USA}
\altaffiltext{14}{School of Information Sciences, University of Illinois, 501 E. Daniel St, Champaign, IL 61820, USA}
\altaffiltext{15}{DIRAC Institute, Department of Astronomy, University of Washington, 
3910 15th Ave NE, Seattle, WA 98195, USA }
\altaffiltext{16}{Department of Astronomy, University of Florida,  211 Bryant Space Science Center, Gainesville, FL 32611-2055, USA }

\begin{abstract}
Turbulence is a key process in many fields of astrophysics. Advances in numerical simulations of fluids over the last several decades have revolutionized our understanding of turbulence and related processes such as star formation and cosmic ray propagation.  However, data from numerical simulations of astrophysical turbulence are often not made public. 
We introduce a new simulation-oriented database for the astronomical community: The Catalogue for Astrophysical Turbulence Simulations (CATS), located at \url{www.mhdturbulence.com}. CATS includes magnetohydrodynamic (MHD) turbulent box simulation data products generated by the public codes  {\sc athena++}, {\sc arepo},  {\sc enzo} and {\sc flash}. CATS also includes several synthetic observation data, such as turbulent HI data cubes.  We also include measured power spectra and 3-point correlation functions from some of these data. We discuss the importance of open source statistical and visualization tools for the analysis of turbulence simulations such as those found in CATS.

\end{abstract}


\section{Introduction}
\label{intro}

Magnetohydrodynamic (MHD) turbulence is now recognized as a vital component of galaxy evolution and of interstellar medium dynamics \citep{Armstrong95,ElmegreenScalo,Lazarian09rev}.
The ability to perform direct numerical simulations of turbulence has been the basis for a revolution in the field of fluid dynamics. Numerical simulations of fluids have been highly influential for our understanding of the physical conditions and statistical properties of MHD turbulence both in astrophysical environments \citep{maclow04,Ballesteros-Paredes07a,Mckee_Ostriker2007,Federrath08a,krumreview2014,beattie2020} and in laboratory experiments \citep{Nornberg2006,Bayliss2007}. 
The emergence of the turbulence paradigm and progress that has been made towards describing it analytically \citep[e.g.][]{GS95,Fleck1996} would not be possible if this paradigm were not validated by numerical simulations (see a monograph by \citealt{berlaz19} and ref. therein). 
 Analytic models are limited to idealized conditions and scenarios while numerical simulations can provide more realistic initial and boundary conditions for turbulent flows, as well as taking into account other relevant and important physics, e.g., gravity, magnetic fields, radiation transport, chemistry, and the complexity of the equation of state of the fluid.

In addition to astrophysically motivated inquiry, basic research on the nature of turbulence and the energy cascade has been advanced significantly by simulations. While the first studies of MHD turbulence done by \citet{Iro64} and \citet{Kra65} were based on a model of isotropic turbulence, later studies began to take into account the anisotropies that the magnetic field induces on the turbulence cascade \citep{Mon81,Mat83,She83,Hig84}. The aforementioned theoretical advances would not have been possible without numerical simulations.  It is through numerical testing \citep[e.g.,][]{Cho2003,Beresnyak05a,kowal2007,Kowal2009,2016JPlPh..82f5301F,Kritsuk09b,2016ApJ...822...11P,2018ApJ...853...96L} of analytic models \citep{GS95,lv99, Boldyrev,lazarian2017} that our modern understanding of MHD turbulence theory has been established.  This in turn, initiated a significant change in our understanding of many key astrophysical processes, e.g. the processes of cosmic ray propagation and acceleration \citep{yan_laz02, laz_yan14}, which were in turn tested with numerical simulations \citep[e.g.,][]{xu_yan13}.

\begin{figure*}[ht]
\centering
\includegraphics[width=.8\textwidth]{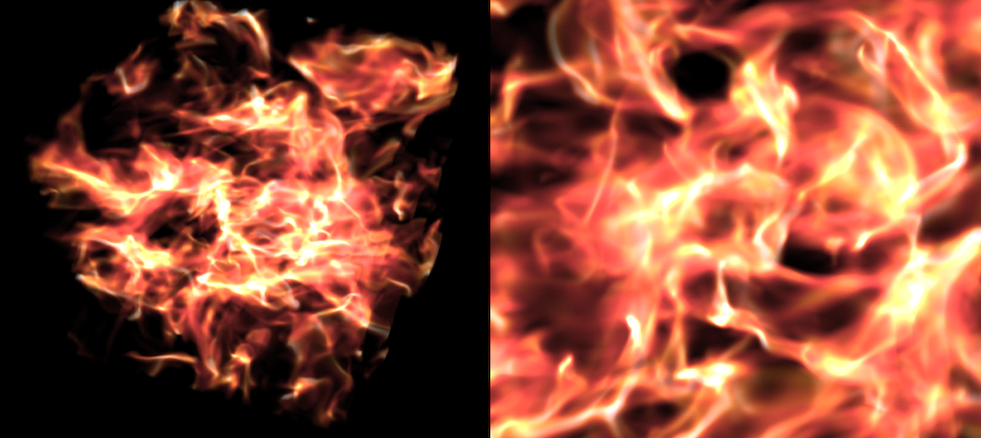}
\caption{Full (left) and closeup (right) views of the data cube from the {\sc Cho-ENO} simulation ``${\rm b1p.01, \mathcal{M}_{\rm s}\sim 7,\mathcal{M}_{\rm A} \sim 0.7}$" generated with \textsc{ytini} as detailed on the \url{www.ytini.com} blog and in the Discussion section of this work.
\label{fig:ytini}
}
\end{figure*} 

Despite the success of numerical simulations for studies of turbulence and related areas of research, there exist very few {\it publicly available simulation data repositories.} 
The most notable open turbulence simulation database is the Johns Hopkins Turbulence Database \citep{Li_2008}; however, this database focuses primarily on incompressible turbulence flows and not on supersonic, highly magnetized, and self-gravitating turbulence such as is found in the interstellar medium (ISM) of galaxies.  Another database, the Starformat Simulations, includes supersonic simulations but is mostly focused on numerical experiments tailored for the star formation problem \footnote{http://starformat.obspm.fr/}. 

In this paper we present a new repository for MHD astrophysical turbulence simulations: The Catalogue for Astrophysical Turbulence Simulations (CATS).  CATS includes MHD turbulence box simulation data products published in previous works from codes including  {\sc athena++}, {\sc arepo},  {\sc enzo} and {\sc flash}.  Additionally, CATS includes synthetic observation data products (e.g., molecular line diagnostic cubes and HI data cubes) and higher-order turbulence statistics, such as the 3-Point Correlation Function (3PCF). CATS can be found at \url{www.mhdturbulence.com}.

This data release paper is organized as follows: In Section~\ref{sims} we give a brief description of the simulations presented as part of the data release and provide the relevant references for the reader to learn more details about the numerical setups. Section~\ref{sims} is organized by code type, as shown in Table~\ref{table.summary}. We also supply the references required for use of these simulations in future published works.  Some of the datasets presented in Section~\ref{sims} also include synthetic observations and/or additional statistical analysis data products.  In Section~\ref{sec:disc}, we discuss and highlight other publicly available tools that are of great benefit to studies involving simulations, including codes for statistics, visualization, and radiative transfer.

\section{Simulations}
\label{sims}

\newcommand{\specialcell}[2][c]{ \begin{tabular}[#1]{@{}l@{}}#2\end{tabular}}
\begin{table*}[t]
\begin{center}
\caption{A brief summary of each simulation suite in the repository.}
\begin{tabular}{l l l l}\label{table.summary}
\\
\hline
Code & Suite Name & Physics & Primary Reference\\
\hline
{\sc Cho-ENO}   & Large Driving      & \specialcell[t]{Isothermal MHD; Driving on large scales}                                        &\specialcell[t]{\cite{Cho2003}, \cite{Burkhart09a} }                      \\
{\sc Cho-ENO}   & Varied Driving     & \specialcell[t]{Isothermal MHD; Driving on various scales}                                     &\specialcell[t]{\cite{bialy2020}}                        \\
{\sc Cho-ENO}   & HI-to-H2 transition& \specialcell[t]{Synthetic HI maps from Isothermal MHD simulations}                           &\specialcell[t]{\cite{bialy2017ApJ...843...92B} }                      \\
{\sc Cho-ENO}   & 3PCF               & \specialcell[t]{3-Point Correlation Function from \\\quad Isothermal MHD simulations}              &\specialcell[t]{\cite{Portillo2018ApJ...862..119P}}    \\
{\sc arepo}     & Core Formation     & \specialcell[t]{Isothermal MHD; Self-Gravity\\\quad Large-Scale Driving}                                &\specialcell[t]{\cite{Mocz2017}, \cite{2018MNRAS.480.3916M}}    \\
{\sc enzo}      & Core Formation     & \specialcell[t]{Isothermal MHD; Self-Gravity\\\quad Large-Scale Driving}                                &\specialcell[t]{\cite{Collins12}}   \\
{\sc enzo}      & Supernova Driving  & \specialcell[t]{Adiabatic HD with ISM cooling\\\quad Supernova driving}                           &\specialcell[t]{\cite{li20a,li20b}}   \\
{\sc flash}     & Multiphase ISM      & \specialcell[t]{Vertically stratified galactic disk section\\\quad ISM cooling; Supernova driving}&\specialcell[t]{\cite{hill2018}}   \\
{\sc flash}     & Zoom-in            & \specialcell[t]{Vertically stratified galactic disk section\\\quad ISM cooling; Supernova driving; Self-Gravity}&\specialcell[t]{\cite{ibanez-mejia2017}, \cite{chira2018b}}   \\ 
{\sc flash}     & Turbulent Box      & \specialcell[t]{Isothermal MHD; Driving on various scales} &\specialcell[t]{\cite{Federrath08a}, \cite{Federrath2010}}   \\
{\sc athena++}  & \specialcell[t]{radiative turbulent\\\quad mixing layer }   & \specialcell[t]{3-D Kelvin-Helmholtz-unstable\\ \quad shear flow with radiative cooling} &\specialcell[t]{\citet{Fielding_2020}, \cite{athena++} }  \\
\hline
\end{tabular}
\end{center}
\end{table*}

Here we detail the data products included in the initial data release of the CATS project.  When using these simulations, please cite this release paper as well as the relevant literature that describes the simulations and data products.

\subsection{{\sc Cho-ENO} MHD Simulations}

We present ten isothermal uniform grid MHD simulations for diffuse ISM applications (i.e., without self-gravity) in Folder \textsc{256}. 
These simulations use a third-order-accurate hybrid essentially non-oscillatory scheme \citep{Cho2003} to solve the ideal MHD equations,
\begin{align}
 \frac{\partial \rho}{\partial t} + \nabla \cdot (\rho \pmb{\varv}) &= 0, \\
 \frac{\partial \rho \pmb{\varv}}{\partial t} + \nabla \cdot \left[ \rho \pmb{\varv} \pmb{\varv} + \left( p + \frac{B^2}{8 \pi} \right) {\bf I} - \frac{1}{4 \pi}{\bf B}{\bf B} \right] &= {\bf f},  \\
 \frac{\partial {\bf B}}{\partial t} - \nabla \times (\pmb{\varv} \times{\bf B}) &= 0. \ 
\end{align}

${\bf B}$ is the magnetic field, $p$ is the gas pressure, and ${\bf I}$ is the identity matrix. These simulations have periodic boundary conditions and an isothermal equation of state $p = c_{\rm s}^2 \rho$, with $c_{\rm s}$ the isothermal sound speed.
For the energy source term $\bf{f}$, we assume a random large-scale solenoidal driving at a wave number $k\approx 2.5$ (i.e.~1/2.5 the box size) and that the driving is continuous. 
The simulations have $256^3$ resolution elements and have been described and used in many previous works
\citep{Cho2003,kowal2007,2014ApJ...785L...1C,osti_22521545,bialy2017ApJ...843...92B,Portillo2018ApJ...862..119P}. We provide the simulation data in FITS format and show example slices in Figure \ref{fig:ytini}.

The primary control parameters of the {\sc Cho-ENO} MHD simulations are the dimensionless sonic Mach number, $\mathcal{M}_{\rm s} \equiv  |\pmb{\varv}|/c_{\rm s}$, and the Alfv\'{e}nic Mach number, $\mathcal{M}_{\rm A} \equiv |\pmb{\varv}|/ \langle \varv_A \rangle$, where $\pmb{\varv}$ is the velocity, $c_{\rm s}$  and $\varv_A$ are respectively the isothermal sound speed and the Alfv\'en speed, and $\langle \cdot \rangle$ denotes averages over the entire simulation box.
A range of sonic Mach numbers is presented (see Table~\ref{tab:params}) for two different regimes of Alfv\'enic Mach number. The simulations are sub-Alfv\'{e}nic with $\mathcal{M}_{\rm A}\approx0.7$ (i.e.,~strong magnetic field) or  super-Alfv\'{e}nic ($\mathcal{M}_{\rm A}=2.0$). The initial Alfv\'en Mach number in the super-Alfv\'enic runs is 7.0; however, after the small-scale dynamo saturates, the final value of $\mathcal{M}_{\rm A}$ is roughly 2. The simulations are non-self-gravitating.  The units of the files  are in code units.  These MHD simulations are scale-free; a physical scale may be determined by the user for the box length and density \citep{Hill2008, McKee10b}.  Rescaling of these simulations requires that the sonic and Alfv\'en Mach numbers remain fixed, but other physical quantities (density, velocity, etc.) may change to physical units.
Higher-resolution runs (up to $2046^3$) are available on request.

When using these simulations and related data products, please cite: \citet[][]{Cho2003,Burkhart09a,Portillo2018ApJ...862..119P,bialy2020}. 

\begin{table}[ht]
\begin{center}
\begin{tabular}{lll}
Folder Name     & $\mathcal{M}_{\rm s}$  &$\mathcal{M}_{\rm A}$  \\
\hline
b.1p1    & 0.7 & 2.0 \\
b.1p.32  & 1.2 & 2.0 \\
b.1p.1   & 2.0 & 2.0 \\
b.1p.032 & 4.0 & 2.0 \\
b.1p.01  & 7.0 & 2.0 \\
b1p1     & 0.7 & 0.7 \\
b1p.32   & 1.2 & 0.7 \\
b1p.1    & 2.0 & 0.7 \\
b1p.01   & 7.0 & 0.7 \\
b1p.032  & 4.0 & 0.7
\end{tabular}
\caption{The box average sonic and Alfv\'enic Mach numbers after three eddy-turnover times and corresponding file names for the {\sc Cho-ENO}  Simulations.}
\label{tab:params}
\end{center}

\end{table}

\subsubsection{{\sc Cho-ENO} MHD Simulations: Machine Learning Applications}

A subset of the  simulations described above have been run to several hundred snapshots in order to provide data for machine learning and neural network studies. These are located in the folder \textit{Machine Learning} on the CATS website.

\subsubsection{{\sc Cho-ENO} MHD Simulations: Varying Driving Scales}
\label{subsub: 256}

The scale-length of the driving process has an important effect on the stability of density structures against gravitational collapse and on the chemical structure of atomic and molecular clouds. 
The turbulence driving scale affects the size-scale (i.e., the size distribution) of density fluctuations, which in turn affects the abundances of chemical species and molecules, e.g., HI, H$_2$, ArH$^+$, OH$^+$, H$_2$O$^+$  \citep{bialy2017ApJ...843...92B,Bialy2019}.
Thus, observations of atomic and molecular abundances may be used to constrain the density structure of the turbulent interstellar medium and the properties of turbulence driving.
However, to do that we need to know how the size-scale of density structures relates to the driving scale of turbulence.

To this end, \citet{bialy2020} have carried out a set of turbulent, isothermal, MHD box simulations, driven on different driving scales,  from large-scale driving of order the box size ($k=2.5$, where $k$ is the inverse length in units of the box size), to small-scale driving of $k=7$, at a maximum resolution of $1024^3$.
They found that the characteristic size-scale of density structure, as measured by the density decorrelation length, $L_{\rm dec}$, is nearly proportional to the turbulence driving scale, with a mean ratio $L_{\rm dec}/L_{\rm drive} = 0.19 \pm 0.10$.
This value is an average value  (and a standard deviation error) for simulations of different sonic Mach numbers, driving length-scales, and  line-of-sight viewing orientations.
When the density field is viewed along the large-scale magnetic field, the structure is more compact and the relation to the driving scale is tighter, with $L_{\rm dec}/L_{\rm drive}= 0.112 \pm 0.024$. 
The $L_{\rm dec}-L_{\rm drive}$ relation is key to connecting the chemistry of the ISM  (and potentially also gravitational instability) to the turbulence driving mechanism.

\begin{figure*}[htbp]
\includegraphics[width=0.9\textwidth]{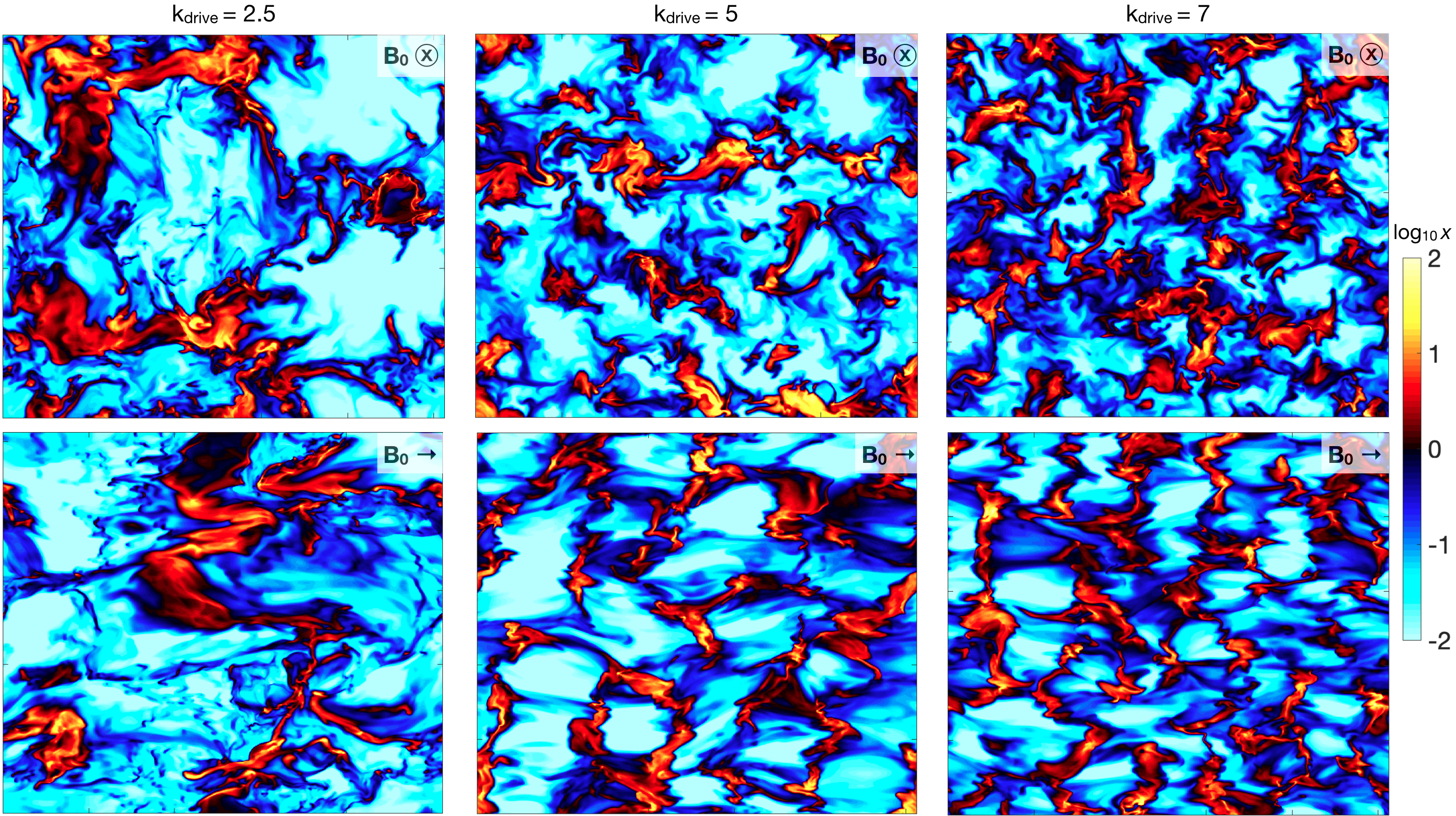}
\caption{Density slices through simulations with different driving scales. The driving scales are at $k=2.5$ (left), $k = 5$ (middle) and $k= 7$ (right). The top panel shows the mean magnetic field along the axis of density integration and the bottom panel shows the density integration perpendicular to the mean magnetic field.  Reproduced with permission from \citet{bialy2020}. 
\label{fig:bialy}
}
\end{figure*}

Folder \textsc{$L_{\rm drive}$}  contains the numerical simulations used in the \citet{bialy2020} study.
These simulations are uniform grid density files (in FITS format) at resolution 256$^3$, similar to those described in the setup in Folder \textsc{256} but
with different driving scales. 
We include simulations with driving scales $k=2.5, k=5$ and $k=7$.  These simulations have sonic Mach number $\mathcal{M}_{\rm s}=7.0$ and Alfv\'enic Mach number $\mathcal{M}_{\rm A}=0.7$.  
We show slices of the simulations with different driving scales in Figure~\ref{fig:bialy}.
Please cite \citet{bialy2020} when using these simulations. Higher-resolution runs (512$^3$ and 1024$^3$) are available on request. 

\subsubsection{{\sc Cho-ENO} MHD Simulations: H{\small I}-to-H$_2$ Transition and Synthetic H{\small I} Observations}
\label{synthetic}

The atomic-to-molecular (HI-to-H$_2$) transition is a key physical process that takes place in the ISM of galaxies \citep[e.g.][]{Savage1977,Gnedin2009, McKee2010, Lee2012, Sternberg14a, Bialy2015_per, Bialy2017_w43}.
The HI-to-H$_2$ transition may be important for the cooling and regulation of star formation (especially in the early universe), and it is also a necessary step for the formation of a wealth of other molecules, e.g., CO, OH, and H$_2$O \citep{Bialy2015_per}, as well as more complex molecules \citep{Tielens2013}.

\citet{bialy2017ApJ...843...92B}  modeled the HI-to-H$_2$ transition in a turbulent medium irradiated by FUV fluxes.
The transition from atomic to molecular form occurs as the FUV radiation is absorbed by dust and also occurs in H$_2$ line excitation (and dissociation). 
The role of turbulence is to create density fluctuations, which in turn have a strong effect on the HI-H$_2$ structure of the gas.
\citet{bialy2017ApJ...843...92B} showed that the 
sonic Mach number and the driving length scale of turbulence directly affect the distribution of the HI column density at the edges of a molecular cloud, where the dispersion of the HI column density distribution increases with increasing Mach number and with increasing turbulence driving scale. Moreover, \citet{10.1111/j.1365-2966.2012.20477.x} investigated the role of the turbulence driving for H$_2$ formation and found that compressive (curl-free) driving of the turbulence is much more efficient at producing molecular gas than solenoidal (divergence-free) driving. This, and the work by \citet{bialy2017ApJ...843...92B}, suggests that observations of HI may potentially be used to constrain turbulence properties in the ISM.

\citet{bialy2017ApJ...843...92B} produced HI
column density maps (units: cm$^{-2}$) for {\sc Cho-ENO} MHD turbulent boxes using a similar setup as presented above by irradiating the numerical box with an isotropic or beamed ultraviolet (UV) field and applying the HI-H$_2$ (atomic to molecular hydrogen) formalism of \citet{Sternberg14a} and \citet{Bialy2016}. This procedure turns the simulation into a
\textit{synthetic observation} that can be used to study HI-H$_2$ physics in the presence of a turbulent medium as well as allowing for comparison with observations.

\begin{figure*}[ht]
\includegraphics[width=.8\textwidth]{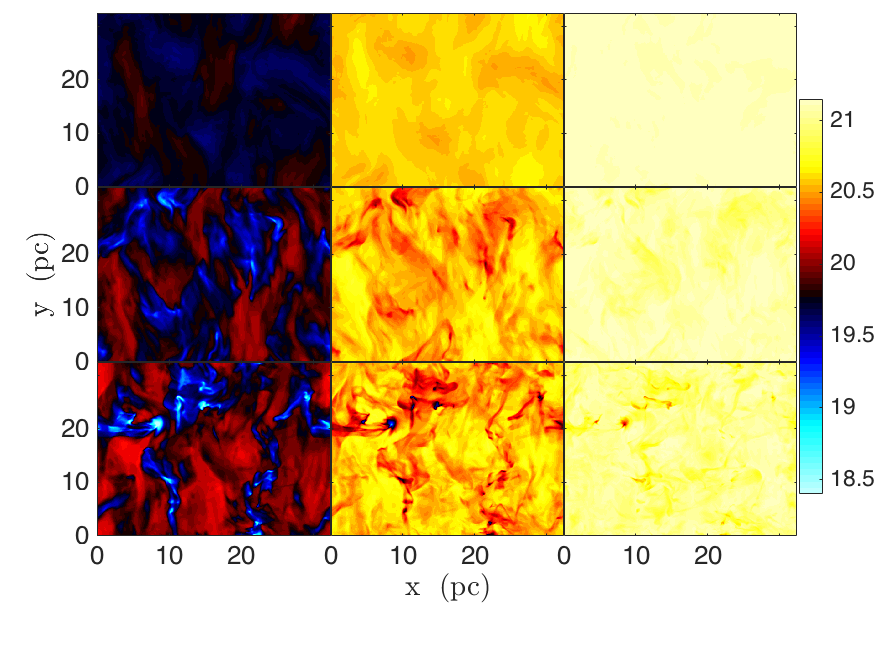}
\caption{Visualization of the synthetic HI column density maps produced from the {\sc Cho-ENO}  simulations. The color bar is $\log_{10}$($N_{\rm HI}$), and the 9~panels correspond to models with $\mathcal{M}_{\rm s}=0.5$, $2$, $4.5$ (from the top row to the bottom row, respectively) and $I_{\rm UV}=0.1$, $1$, $10$ (from the first column to the last column, respectively).
\label{fig:HI}
}
\end{figure*}

The basic setup of these synthetic observations is the following: UV radiation photo-dissociates molecular hydrogen and produces an atomic hydrogen layer at the cloud boundary. The HI column density of this layer depends on the UV intensity, gas density and dust absorption cross section. 
As shown by \citet{bialy2017ApJ...843...92B}, the HI column density also depends on the properties of the turbulence. 
Turbulence produces density fluctuations in the gas, which alter the H$_2$ formation efficiency and self-shielding, thus resulting in
fluctuations in the HI and H$_2$ column density for different lines of sight \citep{bialy2017ApJ...843...92B}.  We provide results for simulations with sonic Mach number $ \Ms = 4.5$ and Alfv\'enic Mach number $ \Ma = 0.7$.
The assumed physical parameters represent those typical to the Milky Way galaxy, with:
1) the UV intensity relative to the \citet{Draine1978} field $I_{\rm UV}=1.3$,
2) the dust absorption cross section in the Lyman-Werner band $\sigma_g = 1.9 \times10^{-21}$ cm$^2$, 
3) a mean density of the cold neutral medium $n = 30$ cm$^{-3}$, and
4) the ratio of the density decorrelation scale to the HI layer scale
$L_{\rm dec}/L_{\rm HI}=0.1$, i.e., on average there are $\sim$ ten density fluctuations along the HI column.
We ask that \citet{bialy2017ApJ...843...92B} please be cited when using these synthetic observations. We show visualizations of the HI column density maps in Figure~\ref{fig:HI}. Additional runs with other parameter sets may be supplied upon request.
In addition, columns and abundances of other molecules (H$_2$, OH$^+$, H$_2$O$^+$, ArH$^+$) are also available upon request. \citep[See][]{Bialy2019}.

\subsubsection{3-point Correlation Function}
\label{sec:3pcf}

\cite{Portillo2018ApJ...862..119P} calculated the 3PCF of {\sc Cho-ENO} MHD turbulent boxes using the Fourier Transform-based approach presented in \cite{Slepian16_alg_WFTs}. The 3PCF is the excess, over and above a spatially random distribution, of density at the vertices of triangles of a given shape and includes phase information that is missed in the 2-Point Correlation Function or power spectrum. For a density field $\delta(\pmb{x})$, the 3PCF, denoted $\zeta$, can be written as a function of two triangle side lengths $r_1$ and $r_2$ and the cosine of the angle between them, $\hat{r}_1\cdot\hat{r}_2$,
\begin{equation}
    \zeta(r_1,r_2;\hat{r_1}\cdot\hat{r_2}) = \left<\delta(\pmb{x})\delta(\pmb{x}+\pmb{r_1})\delta(\pmb{x}+\pmb{r_2})\right>
\end{equation}
where the angle brackets average over all triangles of fixed shape set by $r_1$, $r_2$, and $\theta_{12}$, but with any position $\pmb{x}$ for their vertex. \cite{Slepian16_alg_WFTs} present a fast method to calculate the 3PCF in the Legendre basis where the angle dependence is written in terms of the Legendre polynomials $P_{\ell}$:
\begin{equation}
    \bar{\zeta}(r_1,r_2;\hat{r}_1\cdot\hat{r}_2) = \sum_\ell \bar{\zeta}_\ell(r_1,r_2) P_\ell(\hat{r_1}\cdot\hat{r_2}),
\end{equation}
and the radial dependence is computed on bins described by a binning function $\Phi$:
\begin{equation}
    \bar{\zeta}_\ell(r_1,r_2) = \int_{r \in \Phi(r_1)} r^2 dr \int_{r' \in \Phi(r_2)} r'^2 dr'\; \zeta_\ell(r,r').
\end{equation}
The bars denote that we are obtaining the full 3PCF (and the radial coefficients) binned onto spherical shells. 

We provide 3PCFs calculated up to $\ell_{\rm max}=5$ for $\Nb=32$ radial bins of a constant width of 4~simulation voxels, so that the largest bin probes scales up to 128~voxels, half the simulation box size. On scales larger than half the simulation box we would expect the periodic boundary conditions to cause an order-unity effect, and so the simulation should not be trusted on those scales. There is a 3PCF file for each {\sc Cho-ENO} simulation, with $\delta(\pmb{x})$ being the log of the density fluctuations, i.e., $\delta = [\ln \rho -\left<\ln \rho\right>]/\sigma(\rho)$, where $\sigma(\rho)$ is the density standard deviation. Each file is an $(\ell_{\rm max} +1)\times \Nb \times \Nb$ \textsc{numpy} array representing $\bar{\zeta}_\ell(r_1,r_2)$ with $\ell$ running from 0 to $\ell_{\rm max}$ over the first index and the radial bins running over the other indices. We note that it can be useful to scale $\bar{\zeta}$ by the product of the number of voxels in both radial bins, which scales as roughly $r_1^2 r_2^2$, so we include the number of voxels in all bins as a separate file. Since the cubic grid is discrete, one must actually count the number of voxels in each bin; it is not precisely $r_1^2 r_2^2$ and hence cannot be divided out analytically. Further details are in \cite{Portillo2018ApJ...862..119P}. Please cite \cite{Portillo2018ApJ...862..119P} when using these data products.

\subsection{{\sc arepo}  Moving Mesh Simulations}

We release a simulation suite of solenoidally driven supersonic isothermal magnetized turbulent boxes 
with parameters relevant to the ISM on parsec scales in dense molecular clouds, as simulated in \cite{Mocz2017} and \cite{2018MNRAS.480.3916M} with the {\sc arepo} code. 
{\sc arepo} \citep{2010MNRAS.401..791S} is a moving 
mesh quasi-Lagrangian code which solves the ideal MHD equations with an unstructured vector potential constrained transport solver \citep{1966ITAP...14..302Y,1988ApJ...332..659E,mocz16}  to maintain the 
divergence-free property of the magnetic field. Shocks are captured via 
the use of a Harten-Lax-van Leer Discontinuities (HLLD)  \citep{2005JCoPh.208..315M} Riemann solver. 
In the {\sc arepo} simulations, turbulence is driven in a purely solenoidal (divergence-free) manner in 
Fourier space on the largest spatial scales using an Ornstein-Uhlenbeck 
process 
\citep{Federrath2010,2012MNRAS.423.2558B,2015MNRAS.450.4035F}. 
Snapshots are provided for the properties of the turbulent gas both with and without 
self-gravity switched on. Self-gravity is calculated using a Tree-Particle-Mesh hybrid scheme 
\citep{1995ApJS...98..355X} and leads to the collapse of pre-stellar cores in this physical setup. 

\cite{Mocz2017} ran four isothermal simulations (adaptive $256^3$ resolution) 
with parameters relevant for representing a large region of a giant molecular cloud (GMC), exploring the effect of the (invariant) mean magnetic field $B_0$ threading the system.  
The turbulence is characterized by a sonic Mach number of $\mathcal{M}_{\rm s} \simeq 10$. 
The mean magnetic field (four cases simulated) is characterized by 
the corresponding plasma-beta $\beta_0$ or, equivalently, the Alfv\'enic 
Mach number $\mathcal{M}_{{\rm A},0}$ of the mean-field: $\beta_0= 25, 0.25, 0.028, 0.0025$ or, equivalently, $\mathcal{M}_{{\rm A},0} = 35, 3.5, 1.2, 0.35$. 
The simulations span from very weak seed fields to strong fields whose energy density 
surpasses the turbulent kinetic energy density.
The simulation set establishes quasi-steady state turbulence (indicated by file names with 00), followed by the results obtained when self-gravity is switched on.  Gravity leads to the formation of pre-stellar cores whose collapses are resolved from the GMC parsec scales down to the end of the isothermal collapse (AU scales).  The collapse occurs on the order of the gravitational free-fall time of the cloud. 

The sonic and Alfv\'enic mean-field Mach numbers completely characterize 
the system without gravity. With gravity switched on, the simulations may be
further characterized by a dimensionless virial parameter, $\alpha_{\rm vir} = 5 
v_{\rm rms}^2 (L/2) /(3G M) = 1/2$, where $L$ is the length of the system and $M$ is the total mass.  The virial parameter describes the  strength of the turbulent kinetic energy relative to gravitational potential energy.  The simulations can be appropriately rescaled to different units as long as these dimensionless parameters are held fixed \citep{McKee10b}.

The physical parameters of the 4 simulations (assuming a mass per hydrogen of $1.4~{\rm amu}$) scale as:

\begin{align}
L_0 & = 5.2 \left(\frac{c_{\rm s}}{0.2~{\rm km}~{\rm s}^{-1}}\right)
\left(\frac{n_H}{10^3~{\rm cm}^{-3}}\right)^{-0.5}
\left(\frac{\mathcal{M}_{\rm s}}{10}\right)
 ~{\rm pc} \\
B_0 & = \{ 1.2,12,36,120 \}\times \left(\frac{c_{\rm s}}{0.2~{\rm km}~{\rm s}^{-1}}\right)
\left(\frac{n_H}{10^3~{\rm cm}^{-3}}\right)^{0.5}
 {\rm \mu G} \\
M & = 4860 \left(\frac{c_{\rm s}}{0.2~{\rm km}~{\rm s}^{-1}}\right)^3
\left(\frac{n_H}{10^3~{\rm cm}^{-3}}\right)^{-0.5} 
\left(\frac{\mathcal{M}_{\rm s}}{10}\right)^3 
 M_\odot 
\end{align}
where $L_0$ is the size of the periodic box with a total mass of $M$.

The {\sc arepo} simulations are presented as snapshot HDF5 files that contain fields in the original unstructured Voronoi mesh.  We include full snapshot data of $256^3$ resolution runs. 00 in the file name indicates a run with developed MHD turbulence but no gravity.  Gravity is then switched on for later snapshots.  Two snapshots are included for each run.  
\textbf{We supplementally include re-gridded uniform mesh snapshots as part of the release. 
The CATS AREPO release also includes a sample \textsc{jupyter} notebook that demonstrates how the data can be loaded into {\tt yt}. }At the time of this publication, {\tt yt} does not explicitly support {\sc arepo} simulation data, although that functionality is  expected to be added soon. {\sc arepo} data is read into {\tt yt} as {\sc gadget} particle data, assuming an adaptive smoothing length that scales as the cube root of each Voronoi cell volume.

\begin{figure*}[ht]
\begin{center}
Two representative {\sc arepo} simulations from \cite{Mocz2017}
\begin{tabular}{ccc}
\includegraphics[width=.24\textwidth]{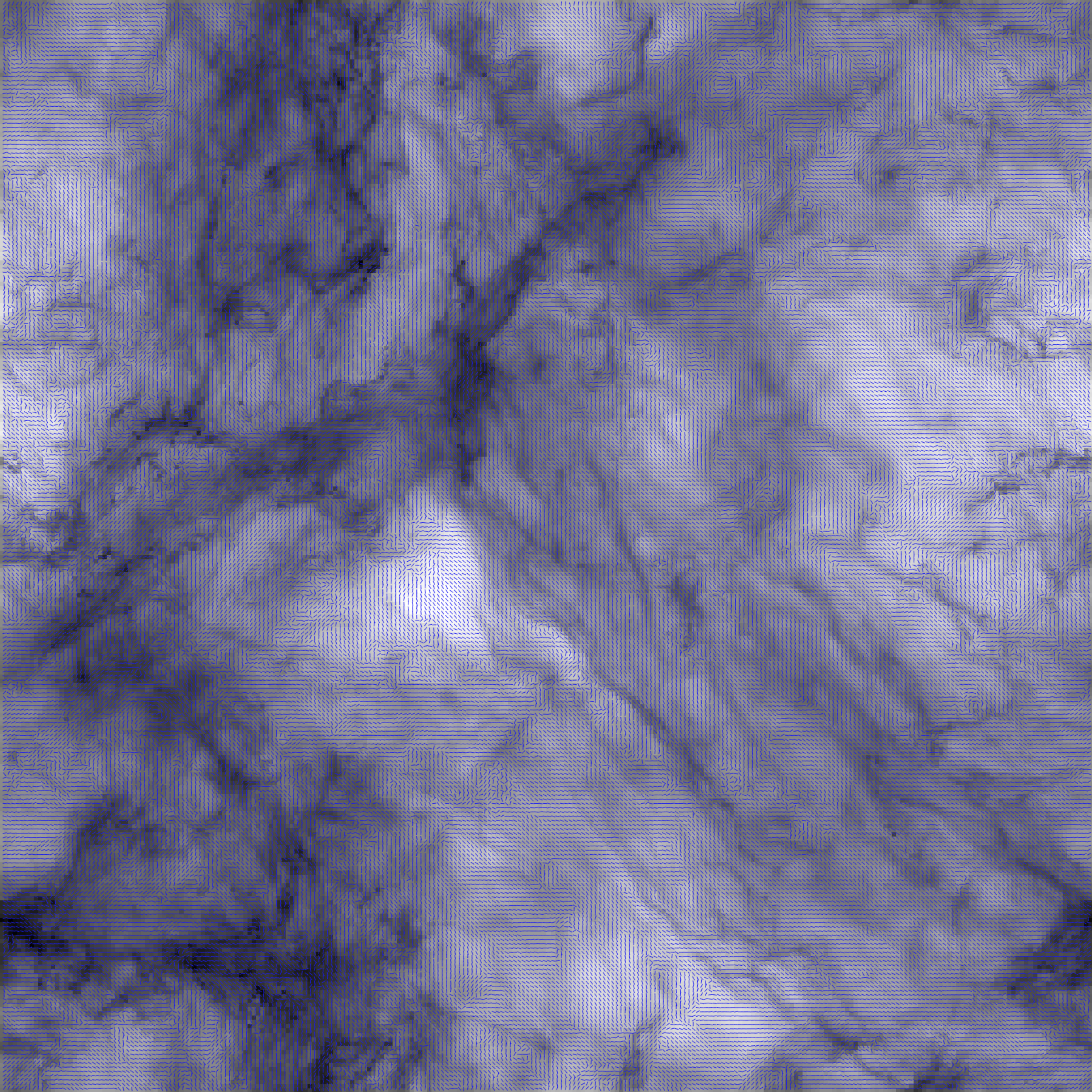} 
\includegraphics[width=.24\textwidth]{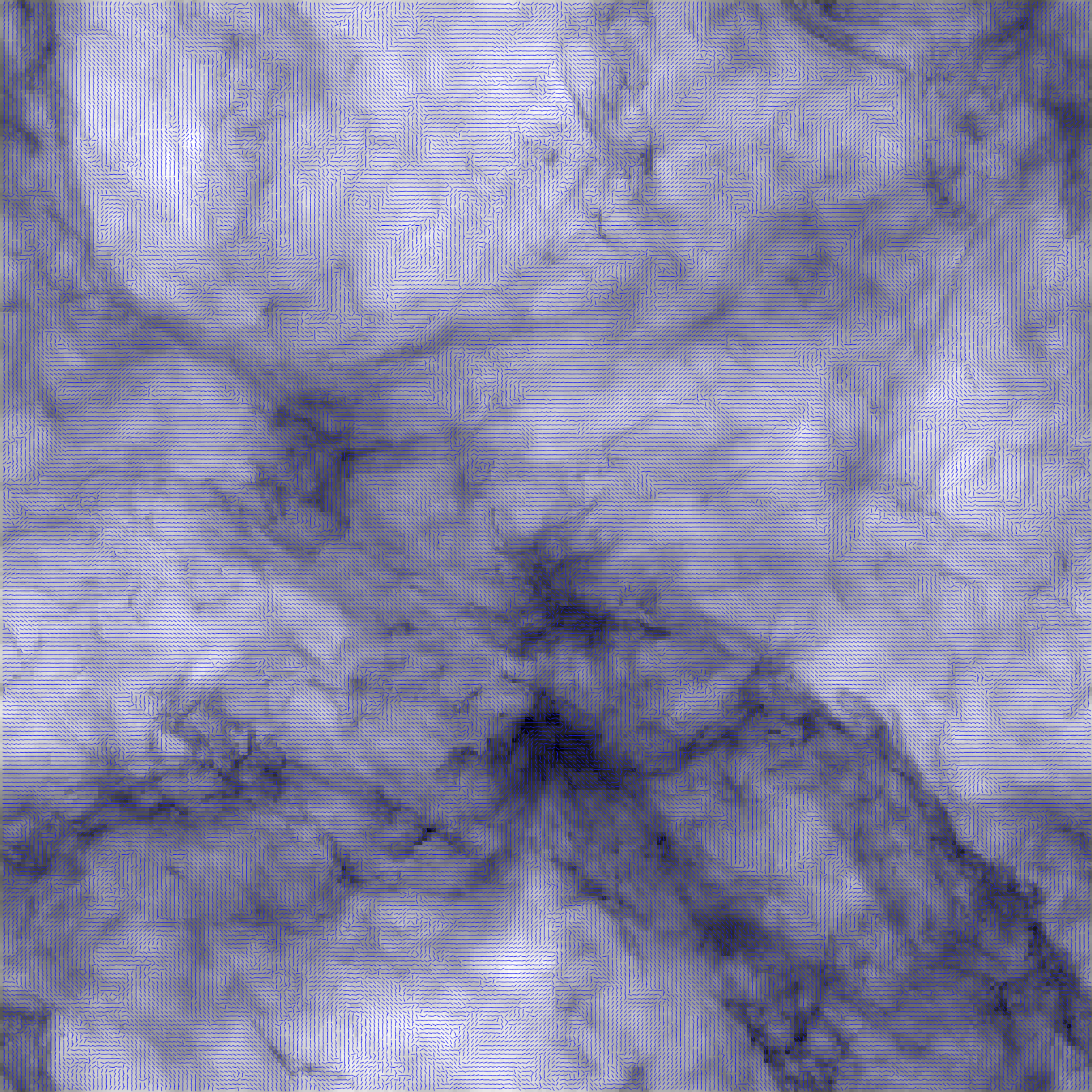} &
&
\includegraphics[width=.24\textwidth]{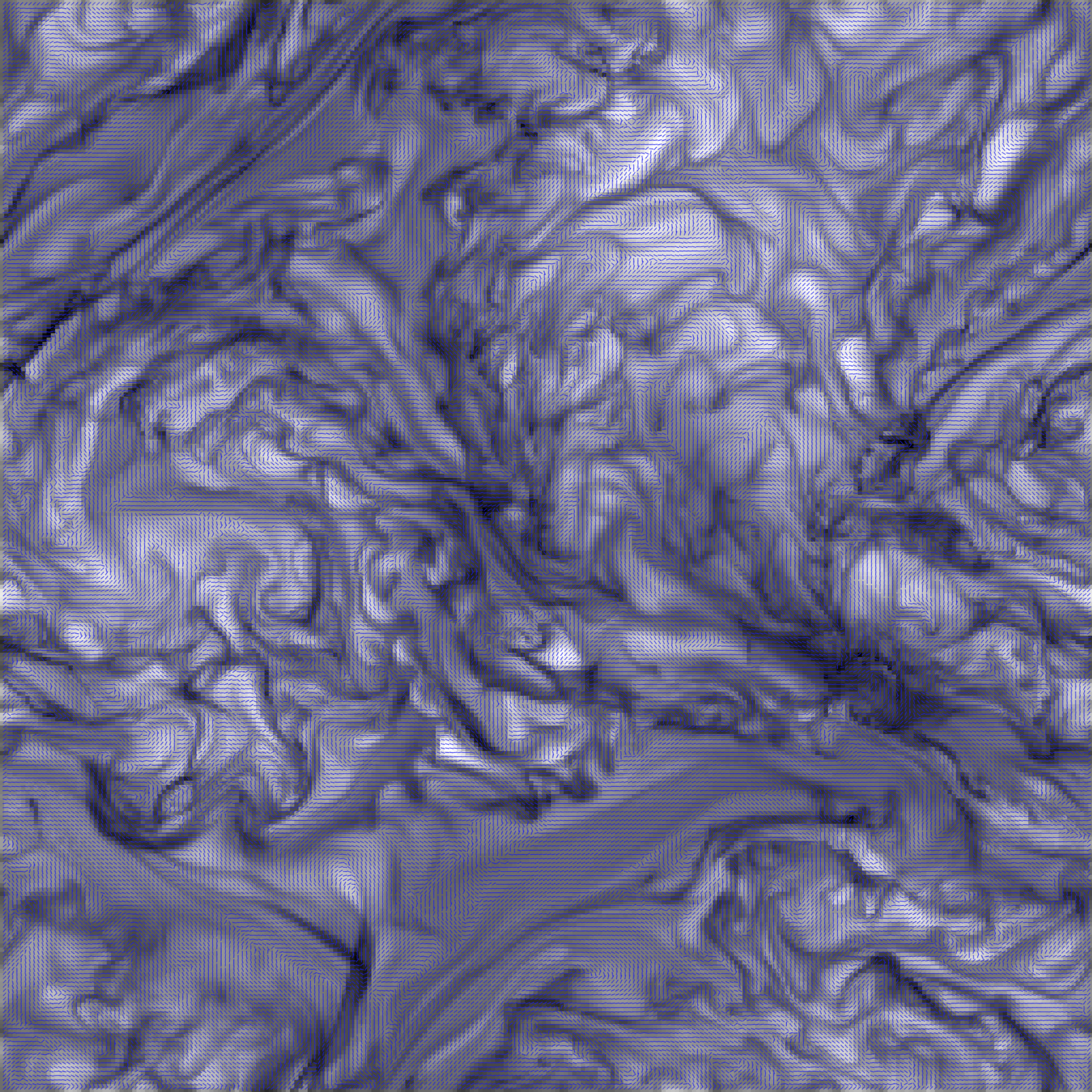} 
\includegraphics[width=.24\textwidth]{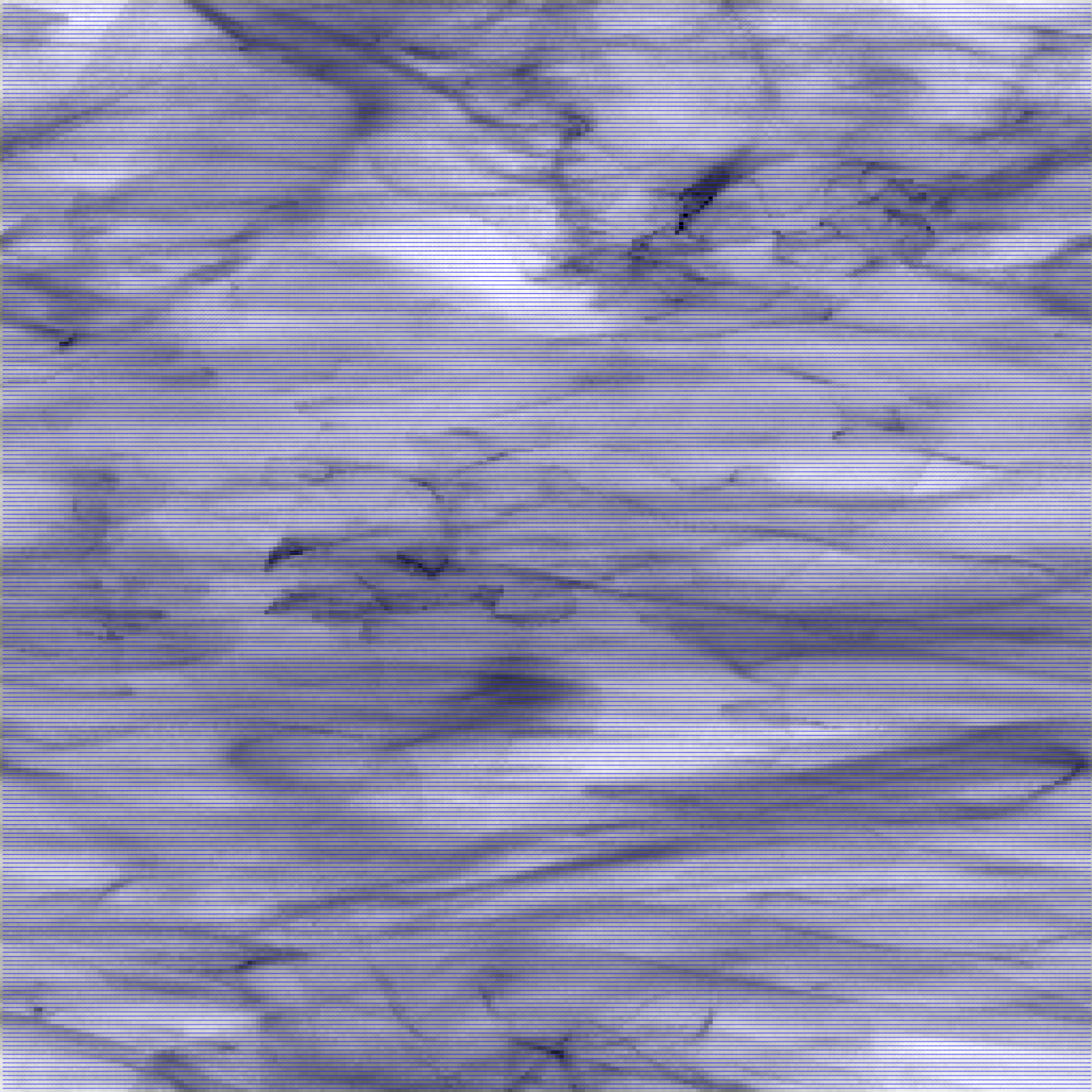} \\
weak mean-field ($\mathcal{M}_{\rm A,0}=35$) & & strong mean-field ($\mathcal{M}_{\rm A,0}=0.35$)  \\
\footnotesize{(projections $\parallel$ \& $\perp$ to mean-field)} & & \\
\end{tabular}
\end{center}
\caption{Visualization of two illustrative cases of {\sc arepo} supersonic isothermal turbulent simulations from  \cite{Mocz2017}.
Shown are density projections perpendicular and parallel to the mean magnetic field along with synthetic UV polarization vectors for a weak magnetic field simulation (\textit{left}) and a strong field simulation (\textit{right}).
Also included in the simulation suite are intermediate field strengths and simulations with self-gravity, which resolve the collapse of pre-stellar cores  (not shown).
\label{fig:arepo}
}
\end{figure*}

Figure~\ref{fig:arepo} highlights two illustrative cases of the simulation suite of \cite{Mocz2017}.

\subsection{ {\sc enzo} Simulations}
\newcommand\jcompphys{J.~Comput.~Phys}
\newcommand{\kkmin}{\ensuremath{k/k_{\rm{min}}}}
\newcommand{\avir}{\ensuremath{\alpha_{\rm{vir}}}}
\def\cs{\ensuremath{c_{\rm{s}}}}
\def\mach{\ensuremath{\mathcal{M}}}
\def\csunits{\ensuremath{c_{\rm{s},2}}}
\def\perrho{\ensuremath{n_{\rm{H},3}}}
\def\betao{\ensuremath{\beta_0}}
\def\rms{\ensuremath{\rm{rms}}}
\def\tff{\ensuremath{t_{\rm{ff}}}}
\def\kms{\ensuremath{\rm{km\ s^{-1}}}}
\newcommand{\msun}{\ensuremath{M_\odot}}
\newcommand{\percc}{\ensuremath{\rm{cm}^{-3}}}

\subsubsection{Driven Turbulent Box Simulations with Self-gravity}

\begin{figure*} 
\begin{center}
\includegraphics[width=0.3\textwidth]{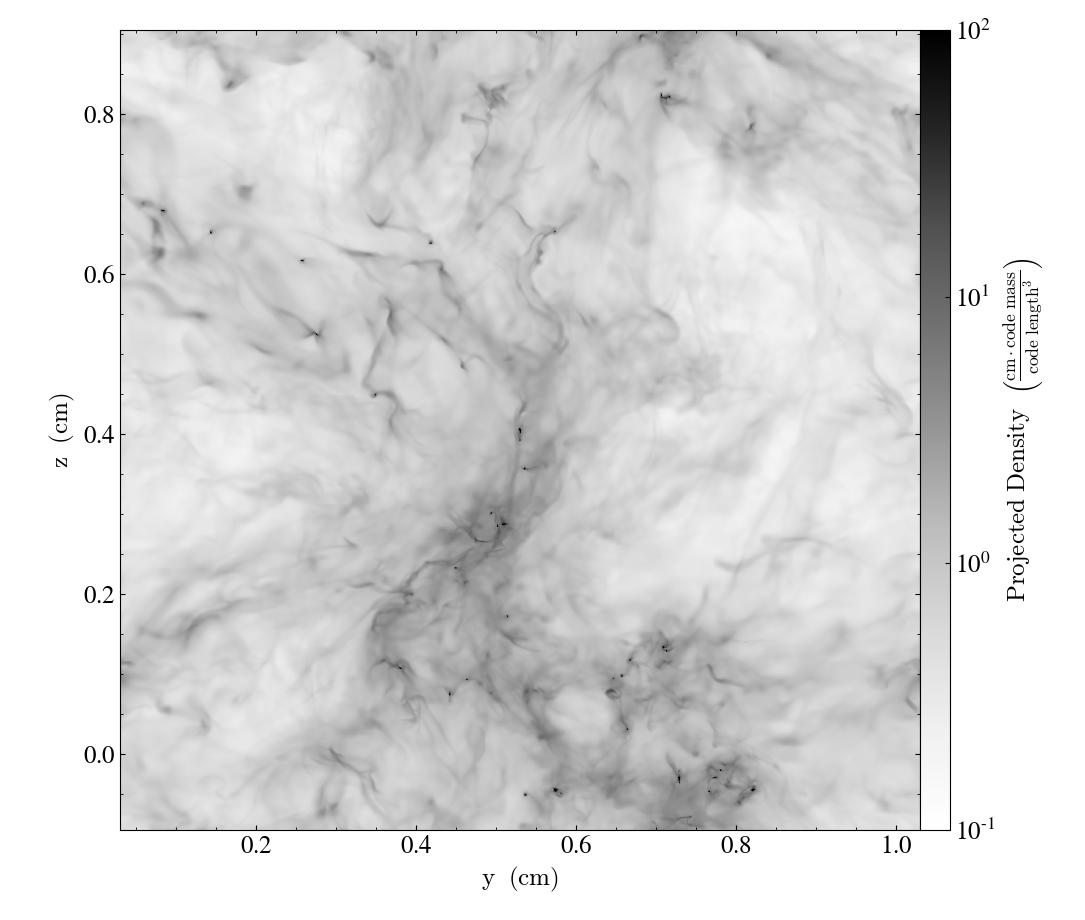}
\includegraphics[width=0.3\textwidth]{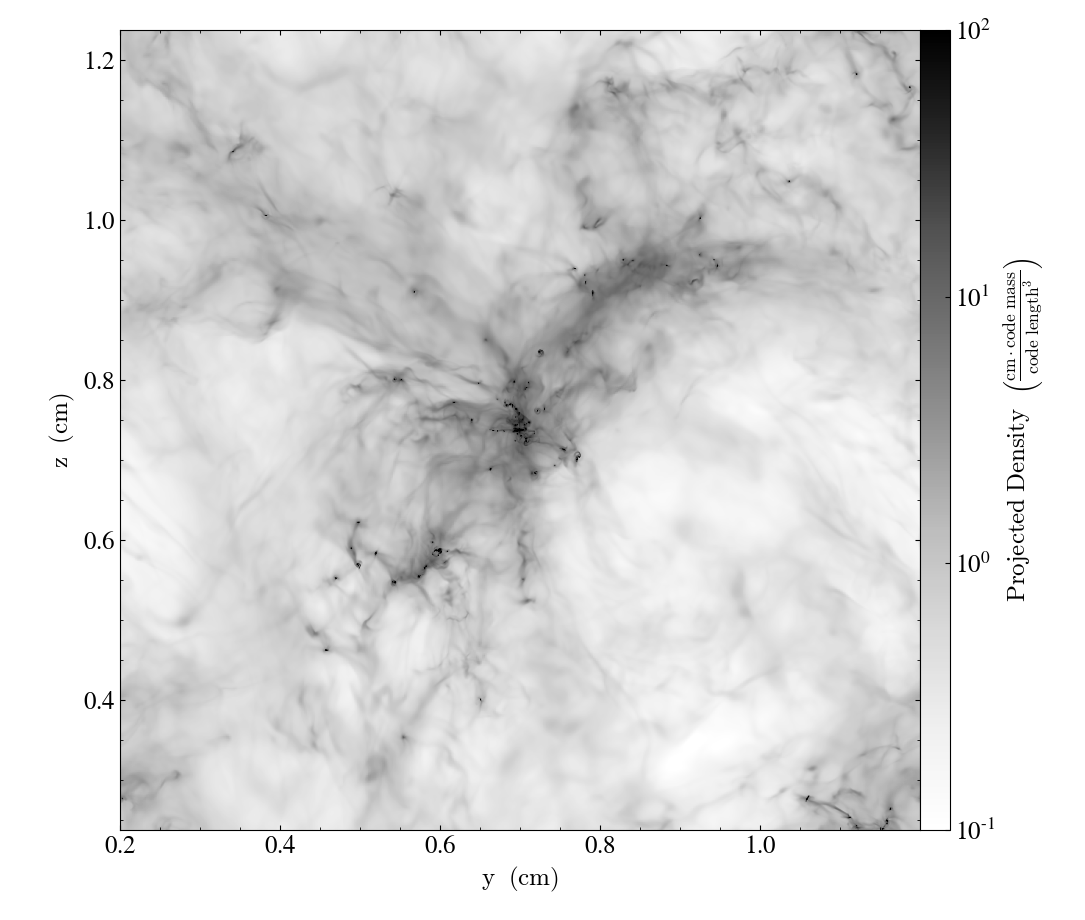}
\includegraphics[width=0.3\textwidth]{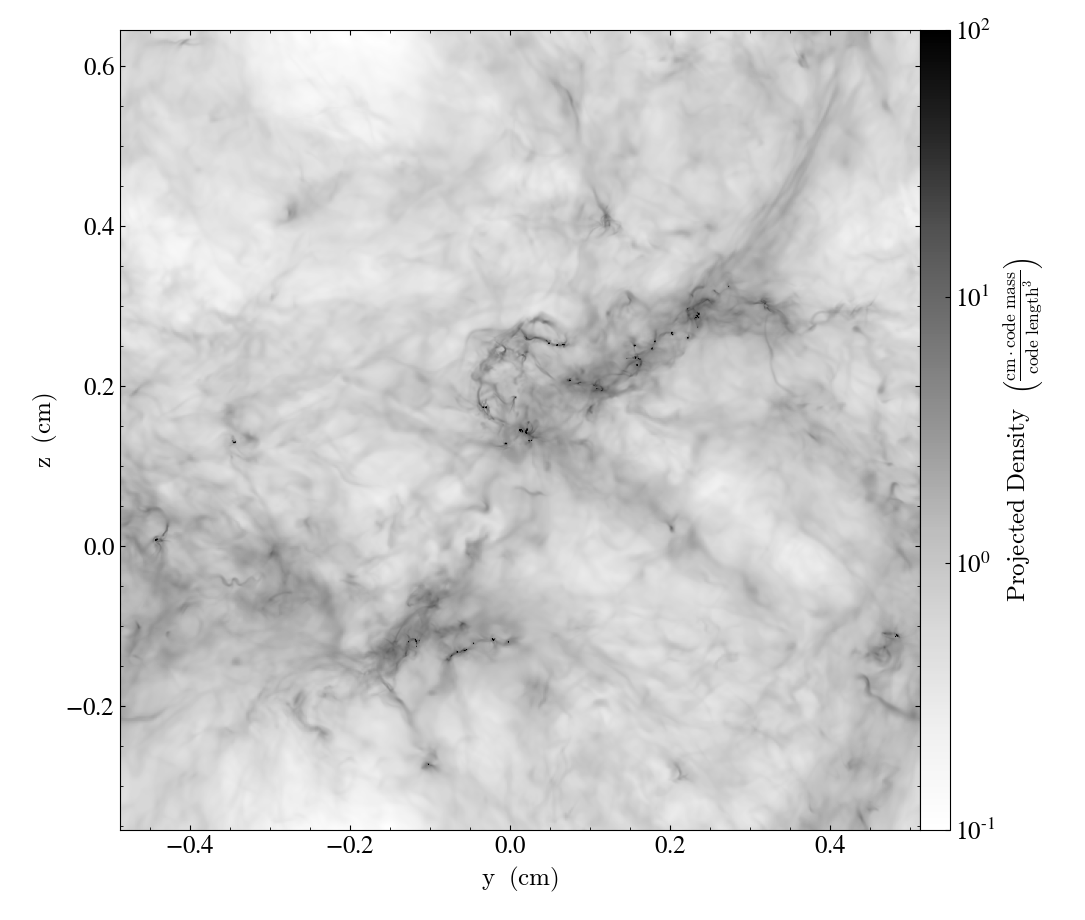}
\caption{ Projections of the density field from the {\sc enzo} core formation simulation.  Left to right, $\beta=0.2, 2, $ and 20, respectively.}
\label{fig.enzocores} 
\end{center} 
\end{figure*}

CATS includes simulations that were first presented in \citet{Collins12}
and employ isothermal, self-gravitating MHD with three values of the mean magnetic field.
Projections of the three simulations can be seen in Figure~\ref{fig.enzocores}.
These datasets used the {\sc enzo} \citep{Bryan14} code, extended to MHD by
\citet{Collins10}.  
This code uses adaptive mesh refinement (AMR)  algorithms developed by \citet{Berger89} and \citet{Balsara01},
the hyperbolic solver of \citet{Li08a}, the isothermal HLLD Riemann solver
developed by \citet{Mignone07},  and the constrained transport method of \citet{Gardiner05} to keep the divergence of the magnetic field to zero.  The
root grid is 512 zones on a side.  Resolution is added adaptively whenever the local
Jeans length, 
 $L_{\rm{J}} = \sqrt{
     \cs^2 \pi/G \rho}$, is resolved by at least 16 zones \citep[which avoids artificial fragmentation and allows us to resolve turbulence on the Jeans scale fairly well; see][]{truelove1997,Federrath_2011}.  This gives an 
 effective
 linear resolution of 8,192 zones.  The simulations were run for $0.6\tff$, where
 the cloud-averaged free-fall time is $\tff= (3 \pi/32 G \rho_0)^{1/2}$. 

The simulations from \citet{Collins12} employ both turbulence and self-gravity with similar Mach numbers as the {\sc arepo} runs. 
 For the {\sc enzo} simulations, we select the  Mach
number, $\mach$, virial parameter, $\avir$, and mean thermal-to-magnetic
pressure ratio, $\betao$, as
\begin{align}
   \mach &=\frac{v_{\rms}}{\cs}= 9,\\
   \avir &= \frac{5 v_{\rms}^2}{3 G \rho_0 L_0^2} = 1,\\
   \betao&=\frac{8 \pi \cs^2 \rho_0}{B_0^2} = 0.2, 2, 20,
\end{align}
where $v_{\rms}$ is the root mean square velocity fluctuation, $\cs$ is the sound speed,
$\rho_0$ is the mean density, $L_0$ is the size of the box, and $B_0$ is the
mean magnetic field. In code units, $\rho_0,$ $L_0$, and $\cs$ are all set to
unity, and the mean magnetic fields are $B_0 = 3.16,\; 1$, and $0.316$.  Note that in code units,
Heaviside-Lorentz units are used, which removes a $4\pi$ from the MHD equations.  Consequently, the magnetic code units differ from cgs units by $\sqrt{4\pi}$ in
addition to the rest of the unit scaling.

We  set the sound speed to be $\cs = 0.2 \kms$, the hydrogen number
density to be $\perrho=n_H/(1000    
 \percc)$ and the mean molecular weight
to be 2.3 amu per particle. The simulations may be re-scaled using the following relations:
\begin{align}
    \tff &= 1.1  \perrho^{-1/2}\; \rm{Myr},\nonumber\\
    L_0 &= 4.6 \csunits \perrho^{-1/2} \;\rm{pc}, \nonumber\\
    v_{\rms} & = 1.8 \csunits\; \kms \nonumber\\
    M &= 5900 \csunits \perrho^{-1/2}\; \msun\nonumber, \\
    B_0 &= \{13, 4.4, 1.3\} \csunits \perrho^{1/2} \;\mu \rm{G}.
\end{align}

The initial conditions for these simulations were generated by a suite of uniform grid
simulations using the piecewise parabolic method on a local stencil (PPML code) \citep{Ustyugov09} without self-gravity. Cubes
with $1024^3$ zones, with initially uniform density and magnetic fields, were
driven solenoidally. Power in the driving was between
wavenumbers $1\leq \kkmin \leq 2$ and driven as in \citet{MacLow99} to maintain the
target Mach number. A full description of this initial turbulent phase can be
found in \citet{Kritsuk09b}. The simulations were then re-gridded to $512^3$ in
a manner that preserved momentum and magnetic flux, and restarted with
self-gravity.  

Three snapshots for each of the three simulations are stored in the CATS
repository, at $t=0.1,\ 0.3$, and $\ 0.6\ \tff$.  The initial $0.1\tff$ is ignored as
the early evolution retains a significant imprint of the higher-resolution initial
conditions, but no substantial effects of self-gravity are seen at this time.
For ease of analysis and data acquisition, fixed resolution cubes at $256^3$ are
stored online. Higher-resolution snapshots, with resolution fixed at $512^3$, and the
full AMR snapshots are available upon request, though these are much larger
files (8~GB and 13~GB, respectively).  The data are stored in code units as
described above.  To simplify units and manipulation, there is a script that
accompanies the datasets that will load the data from the file and, if desired,
create a dataset for use with the \textsc{python}-based code package {\tt yt}. 

When using these simulations in scientific work, please cite \citet{Collins10,Collins12} as well as this release paper.

\subsubsection{Hydrodynamic Simulations of Type Ia Supernova-driven Turbulence in a Hot Medium of Early-type Galaxies}

The simulations provided by \citet{li20a} are also run using the Eulerian code {\sc enzo}. These simulations were set up to investigate how Type Ia SNe (SNe Ia) impact the thermal and dynamical properties of the medium in early-type galaxies. The ISM in these systems is hot and tenuous, with a temperature range of $10^6$--$10^8$ K and a density range of $10^{-3}$--$1$ cm$^{-3}$. The ISM condition is therefore very different from that in  star-forming galaxies. 

The simulations utilize a 3D uniform grid with periodic boundary conditions. The hydro solver is the finite-volume piecewise parabolic method \citep{1984JCoPh..54..174C}. The fiducial runs have $128^3$ cells and the high-resolution runs have $256^3$. The initial condition is a uniform and static medium. SNe Ia exploded randomly in the box at a constant rate. The sizes of the simulation box and the injection zones are chosen based on the fade-away radius of an SN remnant, defined as the radius when the shock wave fades into a sound wave, 
\begin{equation}
\begin{split}
R_{\rm fade} & =  \left(  (\gamma-1) \frac{E_{\rm{SN}}}{4\pi P/3} \right) ^{1/3} \\
& = 48.8~\rm{pc} \ E_{51} ^ {1/3} n_{0.02} ^{-1/3} T_7 ^{-1/3},
\end{split}
\label{eq:rfade}
\end{equation}
where $P$, $n$, and $T$ are the pressure, number density, and temperature, respectively, of the ambient gas, $E_{\rm{SN}}$ is the energy released by an SN, $\gamma$ is the adiabatic index, $T_7 = T/10^7$K, $n_{0.02} = n/0.02 $ cm$^{-3}$, and $E_{\rm{51}} = E_{\rm{SN}}/10^{51}$erg. 

The box length is chosen to be $20\,R_{\rm SN}$. Each SN has a spherical injection zone with a radius  of 0.5$R_{\rm{fade}}$, so the remnant evolution is resolved. The injection is implemented by taking $E_{\rm{SN}}=10^{51}$ erg thermal energy, $m_{\rm{SN}}=$ 1 $\msun$, and a fixed amount of ``color'' (passive scalar) that follows the mass flux. These quantities are evenly distributed in the injection zone. 
Note that $R_{\rm fade}$ depends on the gas pressure, so for runs with different gas pressure the box sizes and injections zones are adjusted according to Equation~\ref{eq:rfade}. The only driving forces for the turbulence are the SNe.

Radiative cooling is included by using a cooling curve \citep{rosen95}. The simulations are run for four cooling times of the initial hot medium. 
A run is uniquely identified by the thermal properties of the medium and the SNe explosion rate.  Gas density, temperature and SN explosion frequencies are varied for different simulation boxes for a parameter study. 
More details on this setup can be found in \citet{li20a}.  

SNe Ia heat the ISM and drive turbulence into the hot medium. SNe feedback in these galaxies is distinguished from that in star-forming galaxies by the fact that the blast waves driven by SNe do not have a cooling phase and therefore do not form a thin shell, but rather fade directly into a sound wave. \citet{li20a} investigated the formation of the cool phase in the hot medium. They found that due to the stochastic nature of the SNe Ia explosions, patches of the hot media not heated by any SNe cool down after an approximate cooling time. Turbulence delays the time of formation of the cool phase. \citet{li20b} examined the thermal structure and turbulence of the hot media under SNe. They have found that the SNe, together with the thermal instability of the hot gas, leave a broad log-normal distribution of density \citep[e.g.,][]{2019MNRAS.482.5233K}.
SNe drive a mild subsonic turbulence. The velocity structure is dominated by compressional, rather than solenoidal, motions. This is different from what occurs in star-forming galaxies \citep[e.g.][]{korpi99,balsara04,kapyla18}. The reason is that in an ISM of an early-type galaxy, the frequency of SNe explosions is low and the interaction among SNe remnants is lacking. As a result, the spherical blast waves are not converted into solenoidal motions.

Furthermore, in an isothermal turbulent medium, there  is a simple correlation between the density fluctuation and the Mach number. The following ratio is generally found to be in the range of 1/3--1 \citep{padoan97,Federrath08a,Federrath2010,konstandin12}: 
\begin{equation}
b  \equiv \frac{\sigma_{\rho,V}}{\bar{\rho}_V}/ \mathcal{M}_{\rm s}, 
\label{eq:b}
\end{equation}
where $\bar{\rho}_V$ is the volume-weighted mean density, $\sigma_{\rho,V}$ is the volume-weighted standard deviation of density, and $\mathcal{M}_{\rm s}$ is the RMS Mach number. 
In contrast, in the hot medium under SNe Ia, this ratio is found to be continuously growing due to thermal instability, reaching 4-20 within a few cooling times.

For {\sc enzo} simulations, the output data can be accessed and analyzed using {\tt yt} . A few example runs are included on the CATS website, each including several snapshots. The naming of the runs indicates the initial condition and the SNe heating rate, e.g., \textit{n0.02\_T1e7K\_H1.02C}
indicates that the initial condition has density $n = 0.02~{\rm cm}^{-3}$, temperature $T =  10^7~{\rm K}$, and SNe heating rate $1.02$ times the cooling rate of the medium. If present, \textit{hr} indicates that the run is a high-resolution run. More data is available upon request. For a full list of the runs, see Table~\ref{table.summary} of \citet{li20a}.

When using these simulations, please cite the following papers: \citet{li20a,li20b} and this release work. 

\begin{figure*} 
\begin{center}
\includegraphics[width=0.48\textwidth]{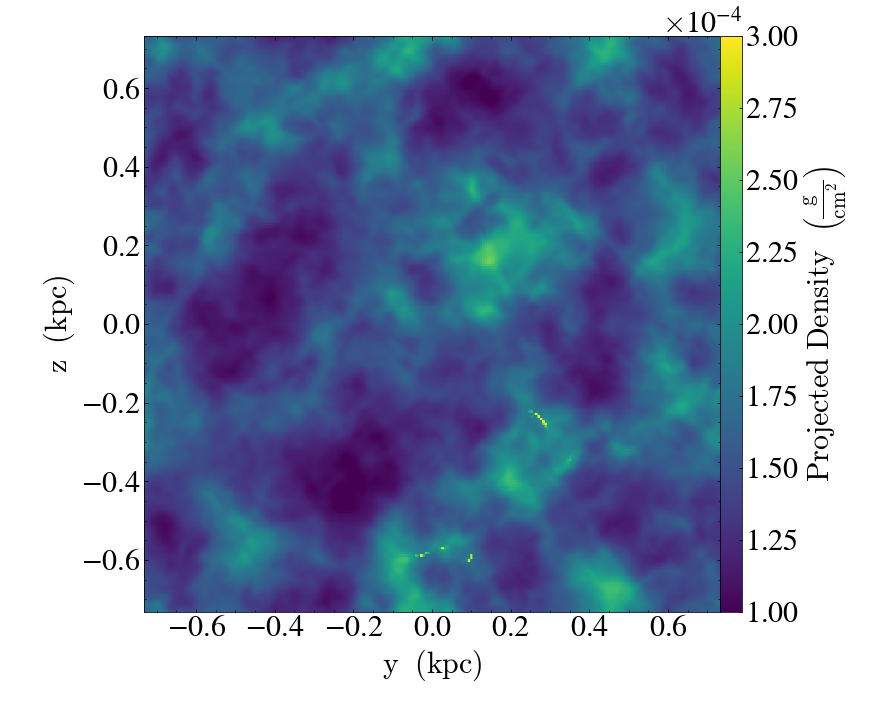}
\includegraphics[width=0.48\textwidth]{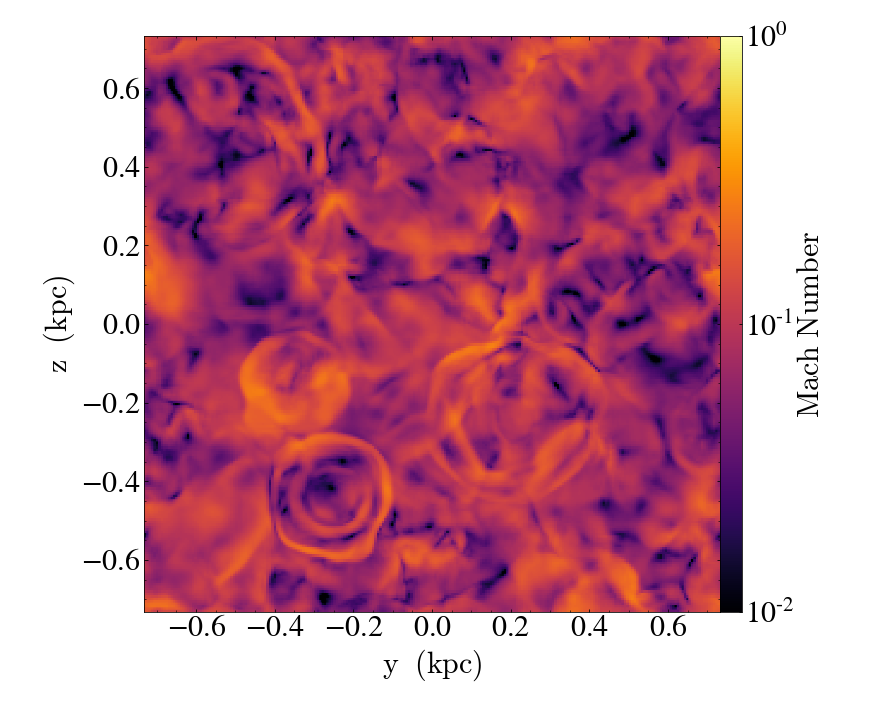}
\caption{Projection of the density and slice of the local Mach number from an exemplar run in \citet{li20a}. }
\label{f:enzo_SNe_Ia} 
\end{center} 
\end{figure*}

\subsection{ {\sc  flash } Simulations}

{\sc flash} is an Eulerian, AMR, MHD code  \citep{Fryxell00a,dubey2012}  that provides a choice of various approximate Riemann solvers for hydrodynamics and MHD.  Available physics modules include self-gravity, external gravity, radiative cooling, diffuse heating from far UV, sink particles for star formation, and turbulent driving.

\subsubsection{ {\sc  flash } Multiphase Interstellar Medium Simulations}
\label{subsub:multiphase}

\begin{figure*}
\plotone{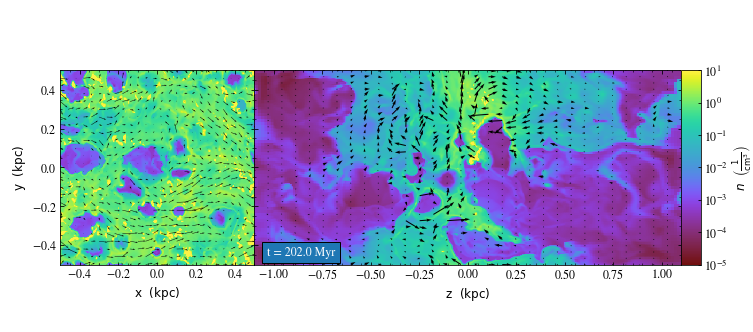}
\caption{Images of number density in a midplane slice (left) and a vertical slice (right) from the {\tt  bx5 pe300  2pc} simulation of \citet{hill2018}. We show magnetic field vectors with  length proportional to $|\vec{B}|$.}
\label{fig:multiphase}
\end{figure*}

\citet{hill2018} used {\sc flash} version~4.2 to run simulations of SN-driven hydrodynamical and MHD turbulence in a stratified, multi-phase section of a galactic disk (neglecting background shear), varying the diffuse heating rate to study the resulting structure of the interstellar medium.  The Riemann solver used is the positivity-preserving, directionally-split HLL3R implementation of \citet{waagan2011}. The models are described fully elsewhere \citep{joung2006,joung2009,hill2012,hill2018}, and we show an image of a slice from one model in Figure~\ref{fig:multiphase}.

These models were run on a rectangular base grid $1 \times 1 \times 40$~kpc in size, centered on the midplane of the Galaxy and extending $|z| = 20$~kpc above and below the midplane. The maximum grid refinement varies with height, reaching values as small as $\Delta x = 0.98$~pc in the innermost 100~pc, and then increasing as shown in Table~2 of \citet{hill2012}.  Periodic boundary conditions were used on the sides, and a zero-gradient outflow boundary condition at the top and bottom, though only a small percentage of the mass ultimately leaves the vertical boundaries.  The gas starts in hydrostatic equilibrium with a static background potential described by \citet{hill2018}, with the innermost kiloparsec above and below the midplane having a temperature $T = 1.15 \times 10^4$~K and gas beyond that being given coronal temperatures of $T = 1.15 \times 10^6$~K.  We assume the gas has a uniform mean mass per hydrogen atom of $\mu = 2.36 \times 10^{-24}$~g, so that the number density $n = \rho / \mu$, where $\rho$ is the mass density of the gas.

These models include a diffuse heating rate $\Gamma(T,z)$, modeling photoelectric heating by far UV photons, and a fixed cooling curve $\Lambda(T)$, modeling collisionally ionized plasma. The balance between these heating and cooling terms establishes a two-phase ISM. The energy density $e$ then evolves as
\begin{equation}
  \partial e_t = n \Gamma(z, T) - n^2 \Lambda(T).
\end{equation}
The value of $\Gamma$ drops off exponentially in the vertical direction with a scale height of 8.5~kpc (high enough that $\Gamma$ varies little within the galactic disk and inner halo). The photoelectric heating term is only applied to gas with temperature $T < 2 \times 10^4$~K, as hotter gas quickly sputters the dust that provides the photoelectrons.  The cooling curve we have chosen results in a multiphase thermal equilibrium for pressures
\begin{equation}
P/k \in ([670, 8300] \mbox{ K cm}^{-3}) (\Gamma / 10^{-25} \mbox{ erg s}^{-1}).
\end{equation}

Multiple populations of SNe are included, each injecting $10^{51}$~erg of energy into 60~M$_\odot$ of gas, along with an approximate treatment of early stellar winds in clusters. SNe are set off at prescribed times based on an average Galactic SN rate with random locations, as the lack of self-gravity and other physical processes important to star formation means we do not meaningfully model star formation. Type Ia SNe are 19\% of the total SN rate, and are given a scale height of $H = 300$~pc; field core-collapse SNe are 32\%, with $H = 90$~pc; and clustered core-collapse SNe are 47\%, also with $H = 90$~pc.  The clustered SNe occur in groups of 7--40 SNe spread over 40~Myr, simulating the evolution of OB associations.  In these locations, we begin by injecting an SN's worth of thermal energy continuously over 5~Myr to model the prompt clearing of gas by stellar winds from OB associations.  SNe and clusters are randomly placed with respect to density concentrations, resulting in clouds living for unphysically long times \citep{hill2018}. Shocks from SNe heat the gas to $\gtrsim 10^6$~K and establish the third phase in the multiphase ISM.

This parameter study focuses on the effect of varying the diffuse far-UV heating rate from 1.09--12.3 $\times 10^{-25}$~erg~s$^{-1}$, as described in Table~1 of \citet{hill2018}, while also varying a small number of other parameters, including the initial magnetic field (either 0 or 5~$\mu$G in the $\hat{x}$-direction), the finest numerical resolution (0.98 to 7.81~pc, corresponding to $1024^3$ to $128^3$ in the plane), the gas surface density (7.5 to 13.4 M$_\odot$~pc$^{-2}$), and the SN rate (17.1 or 34.1 Myr$^{-1}$~kpc$^{-2}$).  In all cases, we began by running low-resolution runs ($\Delta x = 7.81$~pc) for 160~Myr to establish the galactic fountain flow.  We then improved the resolution in steps, reaching $\Delta x =  3.91$~pc at 180~Myr,  $\Delta x = 1.95$~pc at 190~Myr, and, in the run with the highest heating rate, $\Delta x = 0.980$~pc at 190~Myr.  The lower-resolution runs were also continued to 200~Myr to enable resolution studies.

 CATS includes output from the models listed in Table~\ref{table.summary} of \citet{hill2018}, along with example \textsc{python} code to read the HDF5 files using {\tt yt} \citep{Turk11a}. The output files are HDF5 files that include all field variables for each of these runs at a time of $t =200$~Myr, when they have reached dynamical equilibrium, providing models for ISM turbulence in both the plane and the fountain flow of the galaxy. Please cite \citet{hill2018} and this CATS paper if these data are used.

\subsubsection{ {\sc flash } Zoom-in Molecular Cloud Simulations}

\citet{ibanez-mejia2016} took MHD models of multiphase, supernova-driven, stratified turbulence that were identical to those presented in Sect.~\ref{subsub:multiphase} (except as mentioned below), but that included gas self-gravity using the multigrid \citep{huang1999} scheme implemented in {\sc flash} \citep{ricker2008,daley2012}, in addition to a background galactic potential (described in \citet{ibanez-mejia2016} in detail). These models were run with {\sc flash} version~4.2.2. The self-gravity was only enabled after the evolution to a dynamical equilibrium, described in Sect.~\ref{subsub:multiphase}, had completed at $t = 230$~Myr, after the explosion of over 7500~SNe. In a subsequent paper, \citet{ibanez-mejia2017} zoomed in on individual newly forming dense clouds, increasing the maximum numerical resolution within collapsing regions to as high as $\Delta x = 0.06$~pc.  This provides data on the structure and evolution of a dense cloud embedded within a turbulent background having thermal and dynamical structure characteristic of the ISM.  This contrasts with the assumption of many models of either a periodic box or a uniform sphere as the initial condition.

These models focus on ISM conditions at the solar circle, with surface density $\Sigma = 13.7 M_\odot$~pc$^{-2}$ and an SN rate of 34~Myr$^{-1}$~kpc$^{-2}$, with the same distribution of types as in Sect.~\ref{subsub:multiphase}, but a slightly larger scale height for the Type Ia SNe of 325~pc.  They all begin with a uniform magnetic field chosen to have plasma $\beta = 2.5$ everywhere which then evolves under the influence of a local turbulent dynamo driven by the SNe \citep[e.g.][]{balsara2004}.  The mean mass per particle in these models is $\mu = 2.17 \times 10^{-24}$~g, assuming atomic hydrogen with helium fraction of 0.97 and 0.3\% heavier elements.  Superbubbles are given the local bulk gas velocity on formation, with a maximum of 20~km~s$^{-1}$.  (In retrospect, this assumption caused many superbubbles that were initialized in low-density regions to move too quickly, and sometimes even escape into the low halo.)  The diffuse heating rate was calculated with the FUV flux from \citet{Draine1978} of 1.7 times the \citet{habing1968} value, and a flux scale height of 300~pc.

\begin{figure*}[ht]
\centering 
\includegraphics[width=0.9\textwidth]{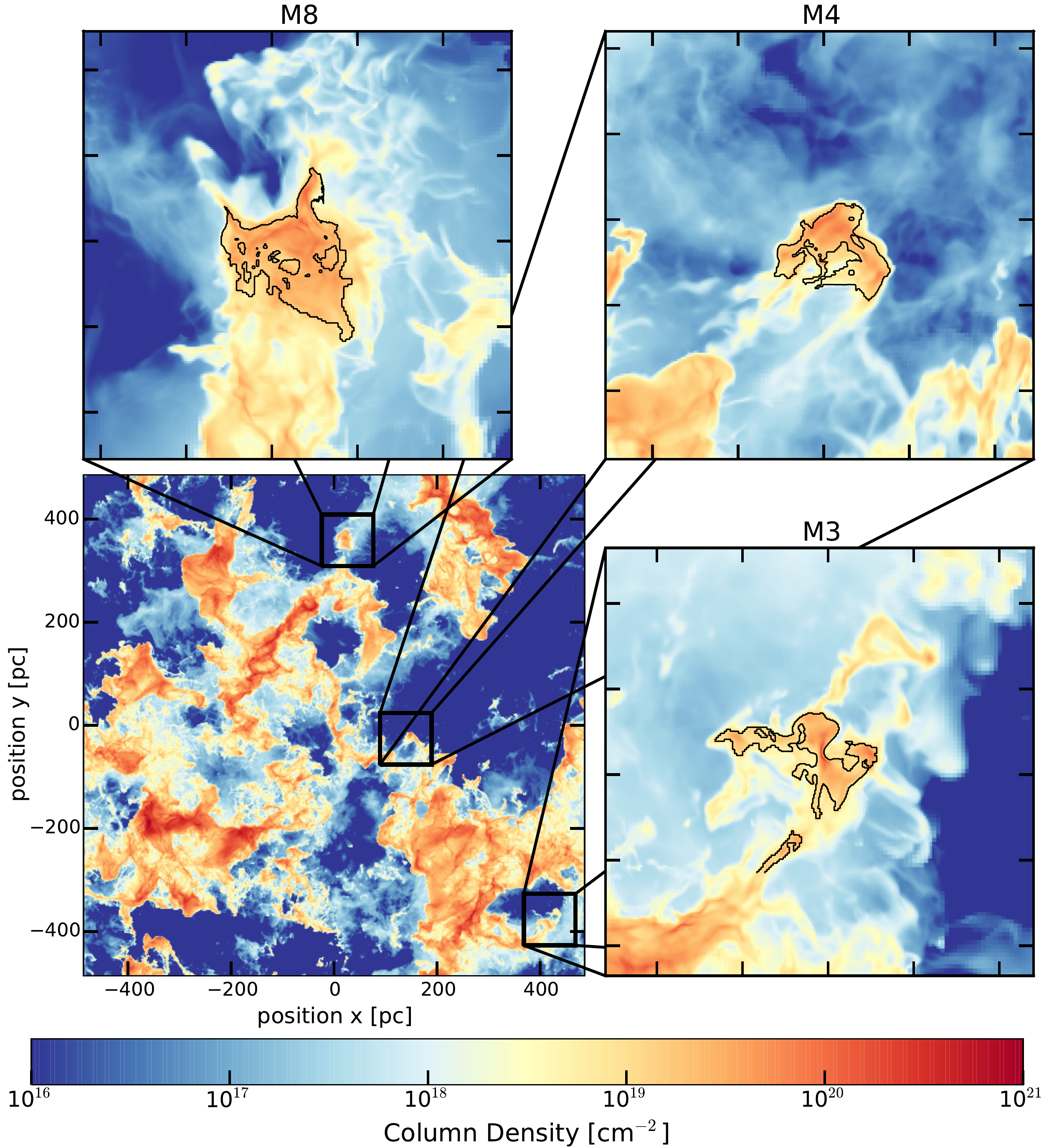} 
\caption{Column density of $\Delta x = 0.47$~pc ISM simulation projected perpendicular to the Galactic midplane at the moment before self-gravity is turned on at $t = 200$~Myr. 
The  (100~pc)$^3$ zoom-in boxes around the target clouds are superposed.  
Close-ups of the zoom-in boxes around each of the clouds are also shown (after about 5~Myr of evolution at higher resolution), with a black contour outlining the $n = 100$~cm$^{-3}$ region. Reproduced with permission from \citet{ibanez-mejia2017}. 
\label{fig:midplane-and-clouds}} 
\end{figure*}

Three clouds were chosen for zoom-in simulations from clouds that started to form earlier in a $\Delta x = 0.47$~pc run than in a $\Delta x = 0.95$~pc run during the last 5~Myr of a 15~Myr extension past $t =230$~Myr. The clouds had initial masses of 3, 4, and 8$\times10^3$~M$_\odot$ at the moment self-gravity was activated, and final masses of a few $10^3$ to $10^4$~M$_\odot$; these are designated M3, M4, and M8 based on their initial mass. The chosen clouds were then re-simulated with $\Delta x = 0.47$~pc within a 100~pc$^3$ box centered on the cloud, while the resolution was reduced for reasons of computational cost to $\Delta x = 1.90$~pc outside in the rest of the midplane (see Table~1 of \citealt{ibanez-mejia2017}). Furthermore, within the zoom-in box, refinement was allowed to continue in dense regions in order to resolve the Jeans length with $n_J =4$ zones \citep{truelove1997}  down to a resolution of $\Delta x = 0.06$--0.12~pc, depending on the mass of the cloud. The runs were ended when more than half the mass of the cloud had exceeded the critical Jeans density at a resolution of $\Delta x$.  Figure~\ref{fig:midplane-and-clouds} shows shows a face-on projection of the box from the simulation with 0.47~pc resolution at the midplane, at the time when the target clouds were identified. The three target clouds are also shown in close-up windows.

The files in CATS are HDF5 files including all field variables from these runs every 0.1~Myr from the start of the zoom-in calculation until the runs were stopped 4.9--6.2~Myr later.  Example \textsc{python} code to read the HDF5 files using {\tt yt} \citep{Turk11a} and convert them to FITS using \textsc{astropy} \citep{astropy2013} is also included.  Please cite \citet{ibanez-mejia2017}, the archival repository \citep{chira2018b}, and this paper if these data are used.

\subsubsection{ {\sc  flash } MHD Turbulence Box Simulations}

CATS includes MHD turbulent box simulations produced with
    {\sc flash}
version~4.6.2. Using  {\sc flash},  we solve the compressible MHD equations on three-dimensional
(3D) periodic grids of fixed side length $L=2$ pc, including turbulence and magnetic fields.
These simulations include a turbulence driving module that produces turbulence driven on large
scales and forced with a large-scale k=1-3 parabolic forcing generator.  We use a mixture of turbulence forcing and include a run that is fully solenoidal
($b=1/3$), fully compressive ($b=1$) and mixed driving ($b=1/2$) using the {\sc StirFromFile} module in {\sc flash} \citep{Federrath08a,Federrath2010}, where $b$ indicates the forcing parameter. The different file names in the online folder show different
values of $b$. We provide snapshots between 2-3 turnover times. 

We include the flash.par parameter file that describes the parameter setup. With the exception of the forcing parameter, all other parameters of the turbulence are the same in these runs. The sonic Mach number is set at $\approx 7.5$ and the Alfv\'enic Mach number is set at $\approx 2$.
The driving velocity amplitude is set to be $v_{\rm drive}=1.5~{\rm km}~{\rm s}^{-1}$ and the sound speed is $c_{\rm s}=0.2~{\rm km}~{\rm s}^{-1}$.
When using these simulations in scientific work, please cite: \citet{Fryxell00a,Federrath08a,Federrath2010,2015MNRAS.450.4035F}, and this release work.

\subsection{{\sc Athena++} Simulations}
\subsubsection{Radiative Mixing Simulations}

\begin{figure*}[ht]
\includegraphics[width=\textwidth]{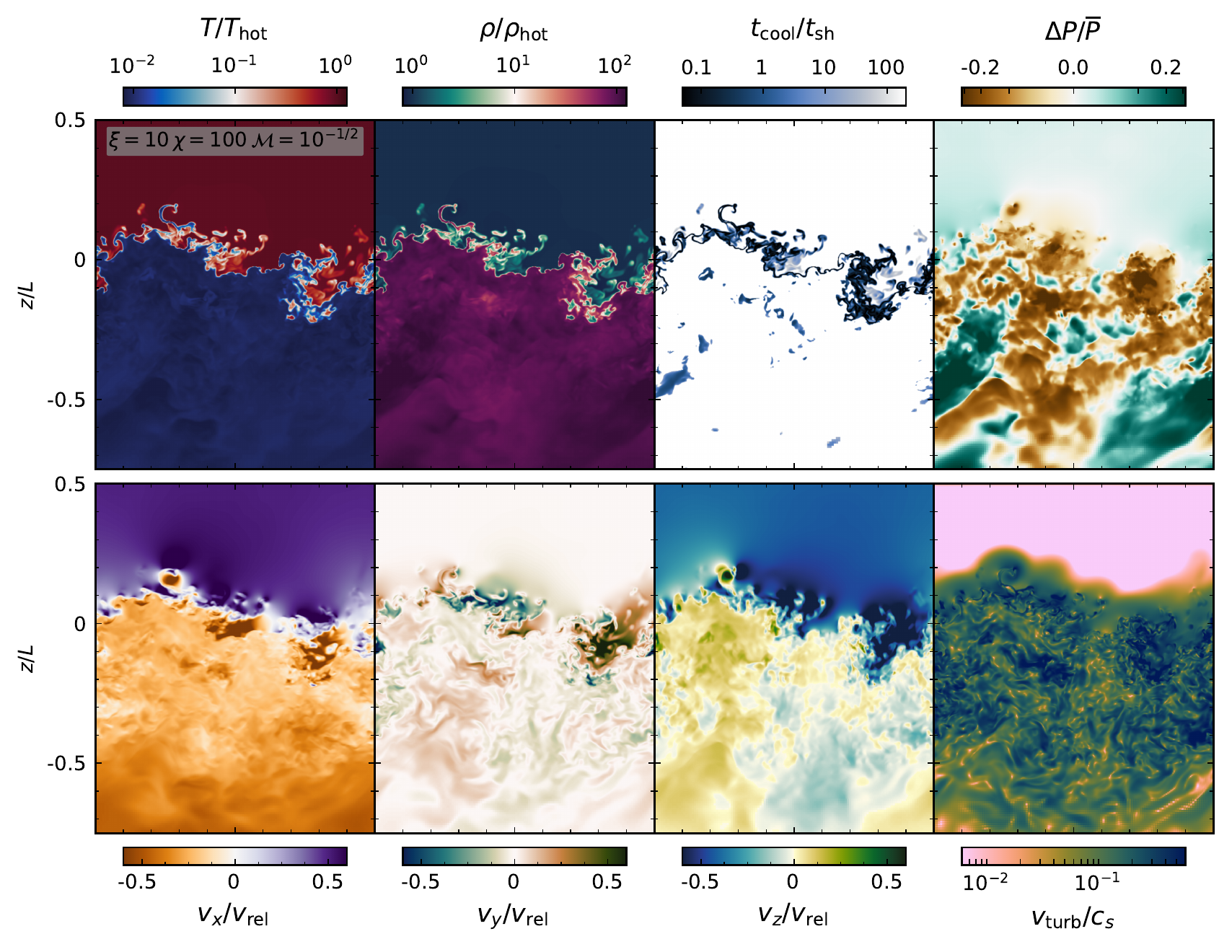} 
\caption{From left to right and top to bottom, slices of temperature ($T$), density ($\rho$), cooling time ($t_{\rm cool}$), pressure deviation ($\Delta P/\overline{P}$), $v_x$, $v_y$, $v_z$, and turbulent Mach number ($v_{\rm turb}/c_{\rm s}$) from the three-dimensional rapidly cooling {\sc athena++} turbulent mixing layer simulation. The background shear flow is in the $\hat{x}$ (horizontal) direction, with the hot gas moving to the right relative to the cold gas. The turbulence, traced by $v_y$, has induced mixing and broadened the shear velocity $v_x$, but the rapid cooling, localized entirely to a thin layer, maintains a sharp gradient between the cold and hot phases. The cooling leads to a flow of the hot gas into the cooling layer, $v_z < 0$. Although the cooling is rapid, there is no signature of the cooling imprinted in the pressure field; instead, the pressure fluctuations correlate with turbulent fluctuations. An animated version of this figure is available \textsf{\href{https://vimeo.com/397632983}{here}}. Reproduced with permission from \citet{Fielding_2020}.}
\label{fig:mixing_layer} 
\end{figure*}

The CATS data release includes two three-dimensional hydrodynamic turbulent mixing layer box simulations run with the {\sc athena++} code framework and described in \citealt{Fielding_2020}. One of the simulations includes strong radiative cooling, while the other does not. An example of the radiatively cooling simulation is shown in Figure~\ref{fig:mixing_layer}. 
These simulations use a fiducial resolution of 128 cells per length ($L$) of the box, stream-wise length of the box and characteristic scale of the mixing layer.  One of the provided simulations has no radiative cooling (in the directory, this run is labeled {\tt tshtcool00}) and the other run has strong cooling, such that the cooling time is ten times shorter than the shear time (in the directory this run is labeled {\tt tshtcool10}). In all other respects, the two simulations are identical. 

The simulation domains are tall skinny boxes that are $L \times L \times 10L$ in shape. The initial conditions of the simulations have hot gas on top ($z>0$) and cold gas on bottom ($z<0$). The gas is in pressure equilibrium. The two phases are moving relative to each other. The relative velocity ($v_{\rm rel}$) of the hot and cold gas is in the $x$-direction and the gradient in the density and velocity is in the $z$-direction. The resolution is highest in the central $L \times L \times 3L$ region, outside of which the resolution rapidly decreases. These simulations have a density contrast of 100 (indicated by the {\tt chi100} label in the directory name) and relative velocity $10^{-0.5} = 0.316$ less than the sound speed of the hot phase (indicated by the  {\tt Mach03} label in the directory name). In code units, the pressure $P = 1$, the relative velocity $v_{\rm rel} = 0.41$, $L = 1$, and the density of the hot phase $n_{\rm hot}=$ 1. The $x$ and $y$ boundary conditions are periodic. The $z$ boundary condition holds the density, pressure, and $x$ velocity constant while maintaining a zero gradient condition for the $y$ and $z$ velocities. There is an initial perturbation to the $z$ velocity that is 4\% of the initial relative velocity. This perturbation has a sinusoidal component with wavelength equal to the box size and a white noise component that is imposed at the grid scale. There is no explicit conduction or viscosity. 

In the radiative simulation, the hydro time-step is constrained to be less than one tenth of the shortest cooling time in the domain. The cooling rate of the material is set such that the cooling time of gas at intermediate temperatures is ten times shorter than the shear time ($t_{\rm sh}/t_{\rm cool} = 10$, where $t_{\rm sh} = L/v_{\rm rel}$). 

The simulation outputs are in HDF5 format. There are 100 outputs per simulation. All of the simulation parameters can be found in the included {\tt athinput.XXX.sh} files. Information regarding the code-specific simulation parameters as well as useful analysis scripts can be found on the {\sc athena++} documentation page. 

More detailed information on these simulations is available upon request or can be found in the primary reference. Visualizations of these simulations can be found  {\sf \href{https://dfielding14.github.io/movies/}{here}}.  When using these simulations in scientific works, please refer to and cite
\citet{Fielding_2020}, the {\sc athena++} code paper \citet{athena++}, and this data release paper. 

\section{Discussion}
\label{sec:disc}

 Shared community resources are increasingly critical for the reproducibility of scientific results and for progress.  This is particularly true of fields that rely heavily on numerical simulations.  Due to proprietary codes and large dataset sizes, numerical data products are often not released to the public. 
In this paper we have presented the CATS data release of astrophysical turbulence simulations, from a variety of codes that are free for public use.

Shared simulation resources such as CATS can benefit both theorists and observers dealing with problems surrounding astrophysical MHD turbulence.
In general, the best strategy for studying a difficult subject like interstellar turbulence is to use a synergistic approach, combining theoretical predictions, numerical simulations, and observational data.
In order to make sensible comparisons between theory, numerics, and astrophysical observations, statistical tools, post-processing of the simulations, and advanced data visualization tools are all necessary.
Below we outline a few additional shared statistical, simulation, and visualization resources open to the community that are compatible with the CATS datasets and have been used in previous MHD turbulence studies.

\subsection{Post-processing Tools: Synthetic Observations}
MHD turbulence simulations provide 3D quantities such as density, magnetic field,  and velocity that are not directly available through observations. Therefore, it is important to realize that simulations must be compared with observational data via the creation of \textit{synthetic observations}.  In practice, this often means a number of post-processing steps to transform the 3D idealized simulation into an \textit{apples-to-apples} comparison with the observation and may include picking a line of sight and applying noise, beam smoothing, and radiative transfer. The output of the post-processing may be a column density map, polarization map, integrated intensity map, spectral line, or position-position-velocity data cube \citep{burkhart13}.  For more applications to the diffuse ISM, 
 turbulence simulations have also explored Synchrotron intensity and Synchrotron polarization fluctuations \citep{Burkhart2012,XuZ16,Kan16,Herron2016ApJ...822...13H,Lazarian2017synchro}.

More specifically, regarding radiative transfer application to the molecular ISM, a number of publicly available radiative transfer codes exist for MHD turbulence simulations.  These include \textsc{RADMC3D}, \textsc{Polaris}, and \despotic. \textsc{RADMC3D} is a radiative transfer code for dust and lines performed using a Monte Carlo method \citep{Dullemond12}. \textsc{Polaris} is a well-established code designated for dust polarization and line radiative transfer in arbitrary astrophysical environments \citep{Reissl_2016,Reissl_2019}.  \despotic~is a library to Derive the Energetics and SPectra of Optically Thick Interstellar Clouds \citep{Krumholz2014MNRAS.437.1662K}.

Here we particularly highlight the \despotic~radiative transfer code, which has been applied to the {\sc enzo} MHD molecular cloud simulations of \citet{Collins12} included in this CATS data release paper.  \despotic's capabilities include calculating spectral line luminosities from clouds of specified physical properties and compositions, along with determining their equilibrium temperatures and chemical compositions. \despotic~is implemented as a \textsc{python} package and its physical model, solutions to the equations it solves, and some tests and example applications are described in \cite{Krumholz2014MNRAS.437.1662K}. Most relevant for CATS, \despotic~can generate look-up tables of emissivity in a variety of molecular lines as a function of gas density and velocity using a range of physical assumptions, e.g., fixing the temperature versus solving for it self-consistently \citep{Onus18a}, or solving for the chemical composition while varying the cosmic ray ionization rate \citep{Armillotta20a}. These tables can then be used together with the analysis package {\tt yt}\footnotetext{\url{www.yt-project.org}} \citep{Turk11a} to generate synthetic position-position-velocity cubes from CATS data.

\subsection{Statistical Analysis of Turbulence}

As mentioned above, the best strategy for studying a difficult subject like interstellar
turbulence is to use a synergistic approach, combining theoretical knowledge, numerical
simulations, and observational data. \textit{ Such a comparison can only be done in a meaningful way with the aid of statistics.} 
While turbulence appears chaotic to the human eye, statistical properties of turbulence averaged temporally or spatially often display reproduciblity and regularity  \citep{Kowal07,Kritsuk07a,Burkhart2015ApJ...811L..28B,Portillo2018ApJ...862..119P}. 

Fortunately, there has been substantial progress
 in the development of techniques to study
turbulence in the last decade. 
Techniques for quantifying aspects of turbulent flows can be tested
empirically using parameter studies of numerical simulations and/or using analytic
predictions.
A statistical view of turbulence allows researchers to describe overall properties of the fluid flow.  For astrophysical turbulence, these properties may include parameters that are important to this data release such as the injection and dissipation  mechanisms,  the  strength  and  properties  of the  magnetic  field  (encapsulated  by  the  Alfv\'enic  Mach number)  and  the  compressibility  of  the  medium (encapsulated  by  the  sonic  Mach  number). 

An example of a more recently developed statistic for studying turbulent flows is the bispectrum and the related three-point correlation function. The bispectrum is the higher order analog to the power spectrum and is the Fourier transform of the 3-Point Correlation Function function (3PCF).  Unlike the power spectrum, the bispectrum includes phase information, which allows  correlations between different spatial frequencies/scales to be explored.  \citet{Burkhart09a} and \citet{Burkhart2016ApJ...827...26B} demonstrated how mode-mode correlations change in different MHD turbulence regimes.
More recently, the 3PCF was applied to similar simulations in \citet{Portillo2018ApJ...862..119P}. They used a fast multipole expansion algorithm introduced by \citet{Slepian15_alg} and extended to use Fourier Transforms (FTs) in  \citet{Slepian16_alg_WFTs}. 
 The 3PCF output of the {\sc Cho-ENO} simulations presented in \citet{Portillo2018ApJ...862..119P} is included in the CATS release (see Section~\ref{sec:3pcf}), as well as ancillary code needed to interpret it.

Previously, there  was  no  common  framework  for  different  turbulence statistics.  Fortunately, a \textsc{python}-based statistics package called \textsc{TurbuStat} was publicly released specifically for application to astrophysical ISM observations \citep{Koch2017,Koch2019}.  \textsc{TurbuStat} implements fourteen observational  diagnostics  of  ISM  turbulence  described  in  the  literature. \textsc{TurbuStat} provides a common framework for running and comparing turbulence diagnostics, including comparisons between simulations and observations.

\subsection{Visualizing Data: ytini}
In additional to performing statistical analysis, it is necessary to visualize simulation data.  The datasets found on the CATS repository are compatible with a number of open source visualization tools.  Excellent visualization packages include \textsc{Glue}  \citep{2015ASPC..495..101B} \footnotetext{\url{http://glueviz.org}} and  {\tt yt}\footnotetext{\url{www.yt-project.org}} \citep{Turk11a}.  {\tt yt} has been cited several times in the above simulation packages  and therefore we will not discuss its many uses further.

\textsc{Glue} is an open-source software aimed at multi-dimensional data visualization and exploration. Equipped with various visualization approaches and selection schemes covering 1D, 2D, and 3D, \textsc{Glue} is distinguished by its linked-view paradigm which enables users to gain insight and understanding from complex models and datasets \citep{2012AN....333..505G}. \textsc{Glue} has hybrid hackable user interfaces with both Graphic User Interface (GUI) and custom-written computer code ability. The GUI provides fluidity with a precision that enables it to perform specific common tasks very easily \citep{2015ASPC..495..101B}.

For simulation renderings and movies, the CATS data can also be used with the \textsc{ytini} package. \textsc{ytini} is a set of interfaces and tutorials designed to incorporate the scientific data-driven capabilities of {\tt yt}  into the industry standard special effects software \textit{Houdini}\footnotetext{\url{www.sidefx.com}} \citep{naiman2017}.  By combining these software packages, an intuitive interface to render volumetric data with access to complex lighting, camera and coloring/shading techniques is provided for scientists to produce movie-quality scientific data visualizations.
Figure~\ref{fig:ytini} shows two rendered images of an example FITS density cube generated with \textsc{ytini}. A movie made by \textsc{ytini} of a CATS simulation may be found on the CATS website. A custom \textsc{python} reader has been created for CATS datasets and is detailed in a blog post on \url{www.ytini.com}.

\section{Conclusions}

In this data release paper we presented the Catalogue for Astrophysical Turbulence Simulations (CATS), a new database for open source compressible MHD turbulence simulations and related statistics.  This data release paper represents our initial contributions, which will remain permanently up on the site.  It is highly likely that the database will evolve over time and include additional contributions and simulations.  For all contributions, a {\sc README} file will be provided to explain the dataset and give examples of analysis and usage.

\acknowledgments
B.B. acknowledges the generous support of the Flatiron Institute Simons Foundation for hosting the CATS database and the support of NASA award 19-ATP19-0020. B.B. especially thanks Nick Carriero, Ian Fish, Andras Pataki, and Dylan Simon for assistance with storing the simulations at the Flatiron Institute. The authors wish to thank the anonymous referee for their comprehensive and insightful report. 
D.C. acknowledges compute resources provided by NSF TRAC allocations TG-AST090110, TG-MCA07S014, and TG-AST140008.
C.F.~acknowledges funding provided by the Australian Research Council (Discovery Project DP170100603 and Future Fellowship FT180100495), the Australia-Germany Joint Research Cooperation Scheme (UA-DAAD), and high-performance computing resources provided by the Leibniz Rechenzentrum, the Gauss Centre for Supercomputing (grants~pr32lo) and the Australian National Computational Infrastructure (grant~ek9) in the framework of the National Computational Merit Allocation Scheme and the ANU Merit Allocation Scheme.
M.R.K. acknowledges support from the Australian Research Council through its Future Fellowship (FT180100375) and Discovery Projects (DP190101258) funding schemes, and from the National Computational Infrastructure (NCI), which is supported by the Australian Government (award jh2). 
A.L. acknowledges the support of the NSF AST 1816234, NASA TCAN 144AAG1967, NASA ATP AAH7546 and the Flatiron Institute.  
J.C.'s work is supported by the National R\&D Program through the National Research Foundation of Korea Grants funded by the Korean Government (NRF2016R1A5A1013277 and NRF-2016R1D1A1B02015014).
P.M. acknowledges support for this work provided by NASA through Einstein Postdoctoral Fellowship grant number PF7-180164 awarded by the \textit{Chandra} X-ray Center, which is operated by the Smithsonian Astrophysical Observatory for NASA under contract NAS8-03060. 
Z.S. was supported during some of the period of this work by NASA through Einstein Postdoctoral Fellowship grant number PF7-180167 awarded by the \textit{Chandra} X-ray Center, which is operated by the Smithsonian Astrophysical Observatory for NASA under contract NAS8-03060. Z.S. also acknowledges support from a Chamberlain Fellowship at Lawrence Berkeley National Laboratory held previously to the Einstein.  
M.-M.M.L. was partly supported by NSF grant AST18-15461, and used computational resources provided by NSF XSEDE through grant number TGMCA99S024.
A.S.H. is supported by a National Sciences and Engineering Research Council of Canada Discovery Grant. A.S.H. and M.-M.M.L. acknowledge support by NASA through grant no. HST-AR-14297 from the Space Telescope Science Institute, which is operated by AURA, Inc., under NASA contract NAS 5-26555, and by NASA ATP grant NNX17AH80G. A.S.H. was partially supported by NSF grant AST-1442650.
The authors acknowledge the use of the following software for this database: 
{\tt yt} \citep{Turk11a}, 
\textsc{flash} \citep{Fryxell00a},
\textsc{arepo} \citep{2010MNRAS.401..791S}, \textsc{athena++} \citep{athena++},
\textsc{SciPy} \citep{SciPy}, 
\textsc{Matplotlib} \citep{hunter2007matplotlib}, 
\textsc{HDF5} \citep{Fortner1998HDF,Koranne2011}, 
\textsc{astropy} \citep{astropy2013}

\bibliography{refsMASTER.bib}

\end{document}